\documentclass{emulateapj}

\def\gsim{\mathrel{\rlap{\lower 4pt \hbox{\hskip 1pt $\sim$}}\raise 1pt
\hbox {$>$}}}
\def\lsim{\mathrel{\rlap{\lower 4pt \hbox{\hskip 1pt $\sim$}}\raise 1pt
\hbox {$<$}}}

\begin{document}

\title{Long-Lasting X-Ray Emission from Type IIb Supernova 2011dh \\
and Mass-Loss History of The Yellow Supergiant Progenitor}  

\author{
Keiichi Maeda\altaffilmark{1,2}, Satoru Katsuda\altaffilmark{3}, Aya Bamba\altaffilmark{4}, Yukikatsu Terada\altaffilmark{5}, Yasushi Fukazawa\altaffilmark{6,7}
}

\altaffiltext{1}{Department of Astronomy, Kyoto University
Kitashirakawa-Oiwake-cho, Sakyo-ku, Kyoto 606-8502, Japan; keiichi.maeda@kusastro.kyoto-u.ac.jp .}
\altaffiltext{2}{Kavli Institute for the Physics and Mathematics of the 
Universe (Kavli-IPMU), University of Tokyo, 
5-1-5 Kashiwanoha, Kashiwa, Chiba 277-8583, Japan}
\altaffiltext{3}{RIKEN (The Institute of Physical and Chemical Research), Nishina Center, 2-1 Hirosawa, Wako, Saitama 351-0198, Japan}
\altaffiltext{4}{Department of Physics and Mathematics, College of Science and Engineering, Aoyama Gakuin University, 5-10-1 Fuchinobe, Chuo-ku, Sagamihara, Kanagawa 252-5258, Japan}
\altaffiltext{5}{Graduate School of Science and Engineering, Saitama University, Shimo-Okubo 255, Sakura, Saitama 338-8570, Japan}
\altaffiltext{6}{Department of Physical Science, Hiroshima University, 1-3-1 Kagamiyama, Higashi-Hiroshima, Hiroshima 739-8526, Japan}
\altaffiltext{7}{Hiroshima Astrophysical Science Center, Hiroshima University, 1-3-1 Kagamiyama, Higashi-Hiroshima, Hiroshima 739-8526, Japan}

\begin{abstract}
Type IIb Supernova (SN) 2011dh, with conclusive detection of an unprecedented Yellow Supergiant (YSG) progenitor, provides an excellent opportunity to deepen our understanding on the massive star evolution in the final centuries toward the SN explosion. In this paper, we report on detection and analyses of thermal X-ray emission from SN IIb 2011dh at $\sim 500$ days after the explosion on {\em Chandra} archival data, providing a solidly derived mass loss rate of an YSG progenitor for the first time. We find that the circumstellar media (CSM) should be dense, more than that expected from a Wolf-Rayet (WR) star by one order of magnitude. The emission is powered by a reverse shock penetrating into an outer envelope, fully consistent with the YSG progenitor but not with a WR progenitor. The density distribution at the outermost ejecta is much steeper than that expected from a compact WR star, and this finding must be taken into account in modeling the early UV/optical emission from SNe IIb. The derived mass loss rate is $\sim 3 \times 10^{-6} M_{\odot}$ yr$^{-1}$ for the mass loss velocity of $\sim 20$ km s$^{-1}$ in the final $\sim 1,300$ years before the explosion. The derived mass loss properties are largely consistent with the standard wind mass loss expected for a giant star. This is not sufficient to be a main driver to expel nearly all the hydrogen envelope. Therefore, the binary interaction, with a huge mass transfer having taken place at $\gsim 1,300$ years before the explosion, is a likely scenario to produce the YSG progenitor. 
\end{abstract}


\section{Introduction}
Evolution of a massive star in the final stage toward the supernova (SN) explosion is one of the main issues in modern stellar astrophysics. Evolutionary paths to SNe IIb/Ib/Ic, sometimes called stripped-envelope (SE-) SNe \citep{filippenko1997}, have been actively debated, since the problem here is directly related to a still-unresolved mass loss mechanism and/or binary evolution. With a progenitor being (nearly) a bare He or C+O star, as analogous to a Wolf-Rayet (WR) star, a question is how the progenitor's envelope has been stripped off for these SE-SNe \citep{nomoto1993,woosley1994,georgy2012,benvenuto2013}. Discovery and confirmation of an unprecedented Yellow Supergiant (YSG) progenitor of SN IIb 2011dh \citep{maund2011,vandyk2011,vandyk2013} brought many questions to the field, but at the same time, it provides a great opportunity to deepen our understanding on stellar evolution toward SE-SNe. With the main-sequence mass of the progenitor estimated as $\sim 12 - 15 M_{\odot}$ \citep{bersten2012}, a single-star evolution model requires a boost in the mass loss rate as compared to the value in the standard prescription \citep{georgy2012}, otherwise it will require a close-binary interaction at some point during the evolution \citep{benvenuto2013}. 

Observationally deriving properties of circumstellar materials (CSM), namely a mass loss rate, should shed light on this issue.  For this purpose (among others), radio and X-ray observations have been actively conducted for nearby SE-SNe. The radio signal is however coupled with still-unresolved relativistic electron acceleration mechanism(s). For SN 2011dh, while a standard mass loss rate of Galactic Wolf-Rayet (WR) stars was {\em assumed} \citep{krauss2012,soderberg2012,horesh2013}, it was also clarified that the radio data (even complemented by optical data) provide only a loose constraint, $4 \lsim A_{*} \lsim 30$ \citep{maeda2012}. Hereafter, $A_{*} \sim (\dot M/10^{-5} M_{\odot} \ {\rm yr}^{-1}) (v_{w}/10^3 \ {\rm km}^{-1})^{-1}$ which is a measure of the CSM density normalized by a typical WR mass loss rate and wind velocity \citep{chevalier2006,chevalier2010}. This is defined by $\rho_{\rm CSM} = 5 \times 10^{11} A_{*} r^{-2}$ g cm$^{-3}$ where the radial position from the SN center (or the progenitor), $r$, is given in cm. 

We note that the radio properties are also related to another issue in the progenitor evolution. The shock velocity inferred from the radio data has been suggested to be one of the measures on the spatial extent of the progenitor's envelope, sometimes referred as `compact' (e.g., WR-like) or `extended' (e.g., RSG or YSG) \citep{chevalier2010}. However, a few recent examples have shown that this is not a perfect indicator of the progenitor's size. The examples include SN 2011dh, which was classified as a compact class from the radio data \citep{soderberg2012}, while the progenitor is identified as an YSG (see above) \citep[see also the case for SN IIb 2011hs: ][]{bufano2014}. Since the radio data provide constraints on the CSM density rather than the progenitor spatial size \citep[except for the very early phase data: ][]{maeda2013b}, this means that a relation between the CSM density (i.e., the mass loss rate and the mass loss velocity) and the progenitor's nature is yet to be clarified. This also provides strong motivation to study the CSM properties around SN 2011dh. 

X-rays provide additionally important information. For SN IIb 2011dh, X-ray observations have been performed in the first month after the explosion, by {\em Chandra}, {\em SWIFT/XRT} \citep{soderberg2012}, and by {\em XMM-Newton} \citep{campana2012,sasaki2012}. However, the interpretation of the X-ray emission mechanism is still controversy for SE-SNe \citep{chevalier2006,maeda2012}. {\em If the X-ray emission would be known} to be thermal, one can conclusively determine the CSM density \citep{fransson1996}, but a problem is that it has not been clarified if the X-ray is thermal or non-thermal for most SE-SNe except for SN IIb 1993J, which is among nearest {\em and} intrinsically brightest X-ray emitter as SE-SNe discovered so far. Generally the low photon statistics does not allow the spectral deconvolution into thermal and non-thermal components \citep[see e.g., ][]{chakraborti2013}, and it is also the case for SN IIb 2011dh \citep{soderberg2012,maeda2012}. For the typical CSM density of $A_{*} \sim 1$, thermal emission is not able to explain the observed X-ray flux level of typical SE-SNe when detected, and thus the inverse Compton (IC) scattering of SN photospheric photons has been sometimes invoked \citep{bjornsson2004,chevalier2006}. This is also the suggestion for SN 2011dh \citep{soderberg2012,maeda2012}, while there might be at least non-negligible thermal emission component as well \citep{sasaki2012}.

In sum, problems in understanding the CSM density, the non-thermal electron acceleration mechanism, and the X-ray production mechanism are all coupled. In this paper, we analyze archival {\em Chandra} data of SN 2011dh at $\sim 500$ days. At this late epoch, the non-thermal processes are ineffective in creating X-ray photons, since the synchrotron emitting electrons should be in the cooling regime and there are not sufficient optical seed photons for the IC emission \citep[e.g., ][]{maeda2013a}. The late-time X-ray emission, if detected, must come from thermal emission either from a forward shock penetrating into the CSM or a reverse shock penetrating into the ejecta. By detecting the thermal X-ray emission from SN 2011dh, we are able to provide a conclusive determination of the mass loss rate. In \S 2, we describe data reduction and spectral analyses where we report on detection of strong late-time X-ray emission. In \S 3, we argue that the late-time X-ray emission is thermal emission from an adiabatic reverse shock penetrating into the outer envelope. In \S 4 (as complemented by discussion in Appendix), we discuss an origin of the early phase X-ray emission, in view of the natures of the reverse shock derived from the late-time emission. The paper is closed with conclusions and discussion in \S 5, where we discuss implications of our findings for the evolution toward the YSG progenitor and for non-thermal mechanism in the early phases. 

\begin{figure}
\begin{center}
        \begin{minipage}[]{0.22\textwidth}
                \epsscale{0.95}
                \plotone{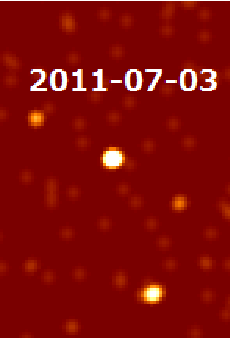}
        \end{minipage}
         \begin{minipage}[]{0.22\textwidth}
                \epsscale{0.95}
                \plotone{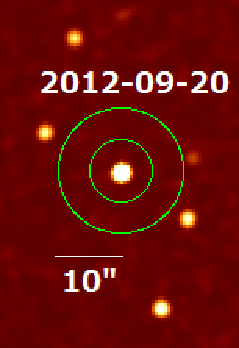}
        \end{minipage}
\end{center}
\caption
{The {\em Chandra} X-ray image of SN 2011dh integrated in 467 - 498 days (right), as compared to the early phase image at 33 days (left). The energy band is 0.3 - 8 keV. The position of the detected source is consistent with the reported position of SN IIb 2011dh within a point spread function. The target spectra are extracted within the inner circular shown by a green circle, and the background is extracted within the annulus defined by the inner and outer circles. 
\label{fig1}}
\end{figure}

\section{Data Reduction and Spectral Analyses}
We analyze the {\em Chandra} archival data of M51, covering the position of SN 2011dh, obtained on 9 September 2012 through 10 October 2012 (obsid: 13812--13816, 15496, 15553, PI: K.D. Kuntz). A total exposure of $\sim 750$ ks has been divided into 7 epochs in 467 - 498 days since the explosion for which we adopt 31 May 2011 \citep{arcavi2011}. In addition, we reanalyze the {\em Chandra} archival data of SN 2011dh in the early phases \citep{soderberg2012} at 12 days (obsid: 12562, PI: D. Pooley, exposure: $\sim 10$ ks) and 33 days since the explosion (obsid: 12668, PI: A.M. Soderberg, exposure: $\sim 10$ ks), as well as the {\em XMM-Newton} archival data \citep{campana2012} at 7 days (obsid: 0677980701, PI: S. Campana, exposure: $\sim 10$ ks). In the following spectral analyses, the error is shown for $1\sigma$ unless otherwise mentioned.

We analyze the data using {\bf CIAO} software (ver. 4.5)\footnote{http://cxc.harvard.edu/ciao/ .} and {\bf HEAsoft} packages (ver. 6.14)\footnote{http://heasarc.gsfc.nasa.gov/docs/software/lheasoft/ .}. An X-ray source is clearly detected at the position consistent with that of SN 2011dh, (R.A., dec.) = (13:30:05.124, +47:10:11.301), within a point spread function (Fig. 1). The position is also consistent with the X-ray source detected in the early phase and there is no doubt that it is SN 2011dh; no such a source was detected in the pre-SN archival images \citep{soderberg2012}. A background-subtracted count rate ($0.3-8$ keV) in each exposure during 467 -- 498 days is $\sim 0.0010 - 0.0015$ counts s$^{-1}$ with the error in the range of $\sim 0.00009 - 0.00020$ counts s$^{-1}$ (within an aperture of $\sim 5"$). Our analysis of the early-phase data results in $0.01284 \pm 0.00117$ counts s$^{-1}$ (12 days) and $0.00464 \pm 0.00070$ counts s$^{-1}$ (33 days), being consistent with the previous analysis \citep{soderberg2012}. Namely, SN 2011dh has faded (only) by a factor of $\sim 4$ from 33 days to $\sim 500$ days. 

\begin{figure}
\begin{center}
\hspace{-1.5cm}
        \begin{minipage}[]{0.45\textwidth}
                \epsscale{0.95}
                \plotone{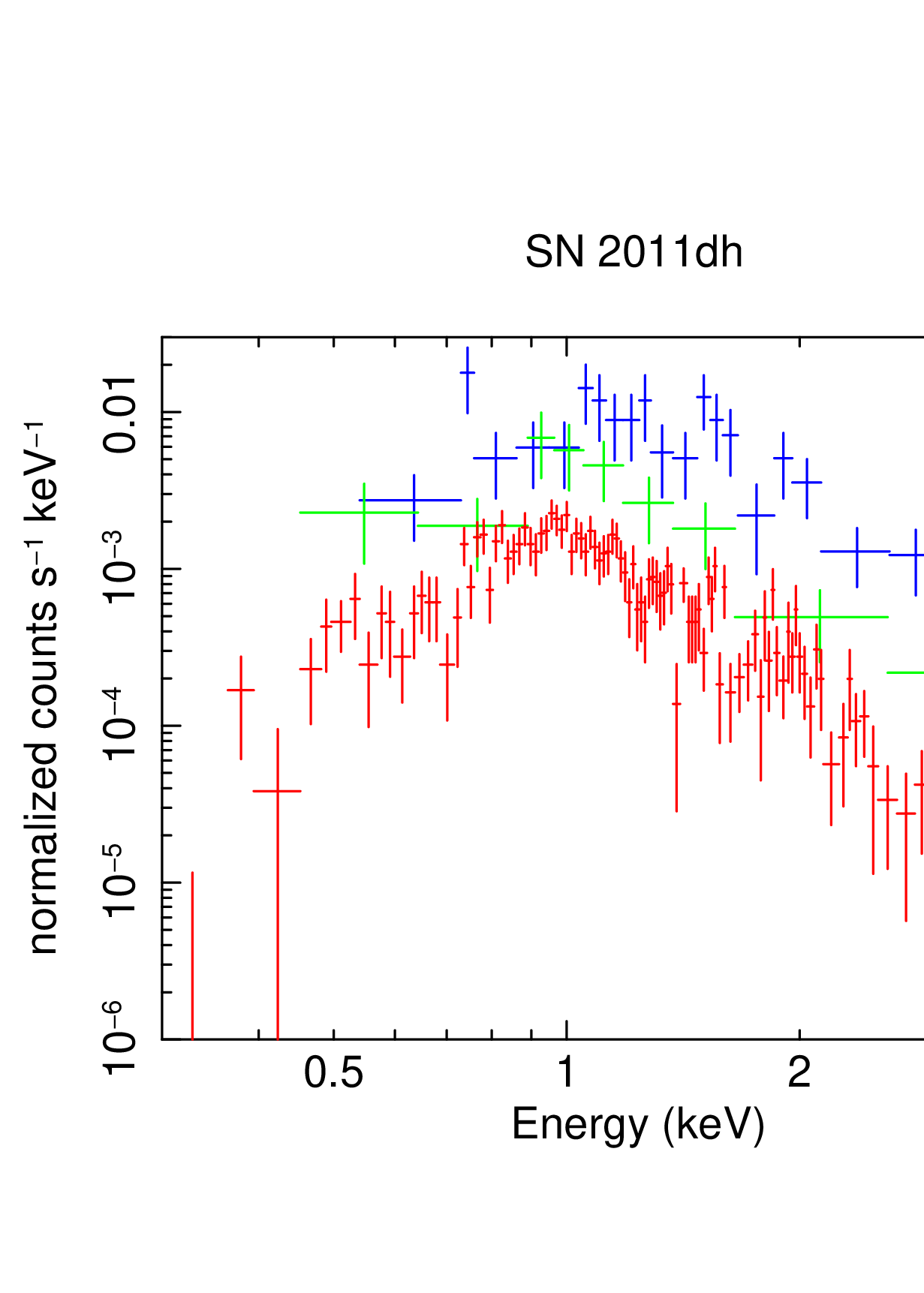}
        \end{minipage}
\end{center}
\vspace{-1.5cm}
\caption
{The extracted {\em Chandra} spectra at 12 days (blue), 33 days (green), and the one integrated in 467 - 498 days (red). 
\label{fig2}}
\end{figure}

Spectra are extracted within a circular region with a radius of 5". Background is extracted within an annulus surrounding the source (5" to 10"), which is free from contaminating sources. The background flux turns out to be negligible (See Fig. 1). We co-add the spectra during 467 -- 498 days in order to create an integrated spectrum with a total exposure of $\sim 750$ ks. The integrated data have the background-subtracted count rate of $0.00134 \pm 4.66 \times 10^{-5}$ counts s$^{-1}$ and the total counts of 1165 in $0.3 - 8$ keV. 
The extracted integrated spectrum is shown in Figure 2, along with the spectra at the early phases (see below). 

We first perform the {\bf Xspec} \citep[ver. 12.8.1.;][]{arnaud1996} model fit to this integrated spectrum. No significant spectral evolution is expected during these epochs, and the integrated spectrum with a high Signal-to-Noise (S/N) ratio allows us to extract detailed spectral information. Our model fit to the integrated spectrum is summarized in Table 1 and Figure 3. We first try an absorbed power law (PL) model and then an absorbed thermal plasma model (either {\bf apec} or {\bf vapec}), and finally an absorbed, two-component thermal plasma model ({\bf apec+apec}). The absorbing column density is first treated as a free parameter, but if it is found to be consistent with the Galactic value \citep[$N_{\rm H} \sim 1.8 \times 10^{20}$ cm$^{-2}$; ][]{kalberla2005}, we alternatively fix this to the Galactic value. For the thermal plasma model, we also separately try fits with the abundance treated as a free parameter or fixed to the solar.  For the solar abundance in the thermal models, we adopt the values from \citet{lodders2003}. 

\begin{figure}
\begin{center}
\hspace{-1.5cm}
\vspace{-3.0cm}
        \begin{minipage}[]{0.45\textwidth}
                \epsscale{0.9}
                \plotone{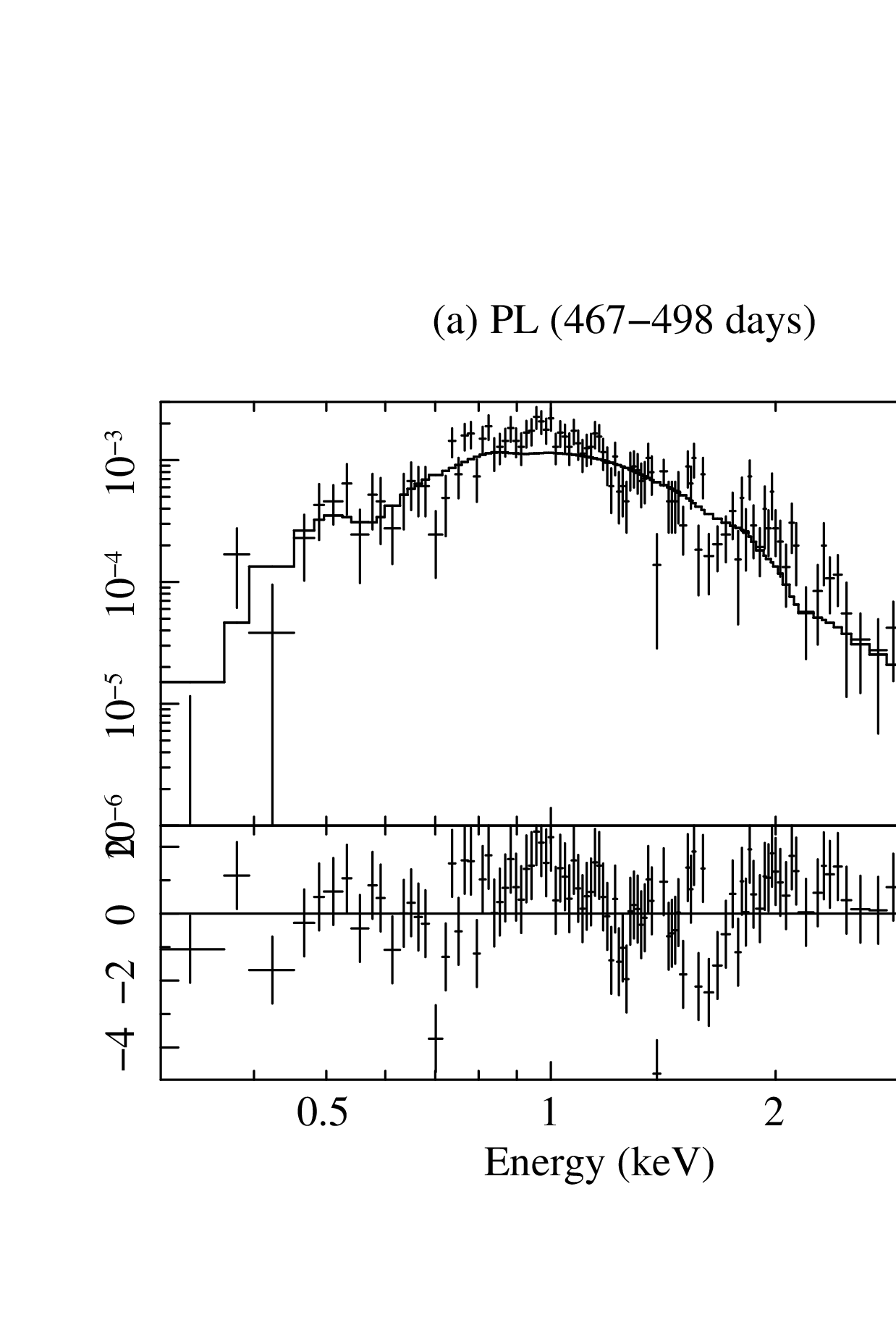}
        \end{minipage}\\
\hspace{-1.5cm}
\vspace{-3cm}
        \begin{minipage}[]{0.45\textwidth}
                \epsscale{0.9}
                \plotone{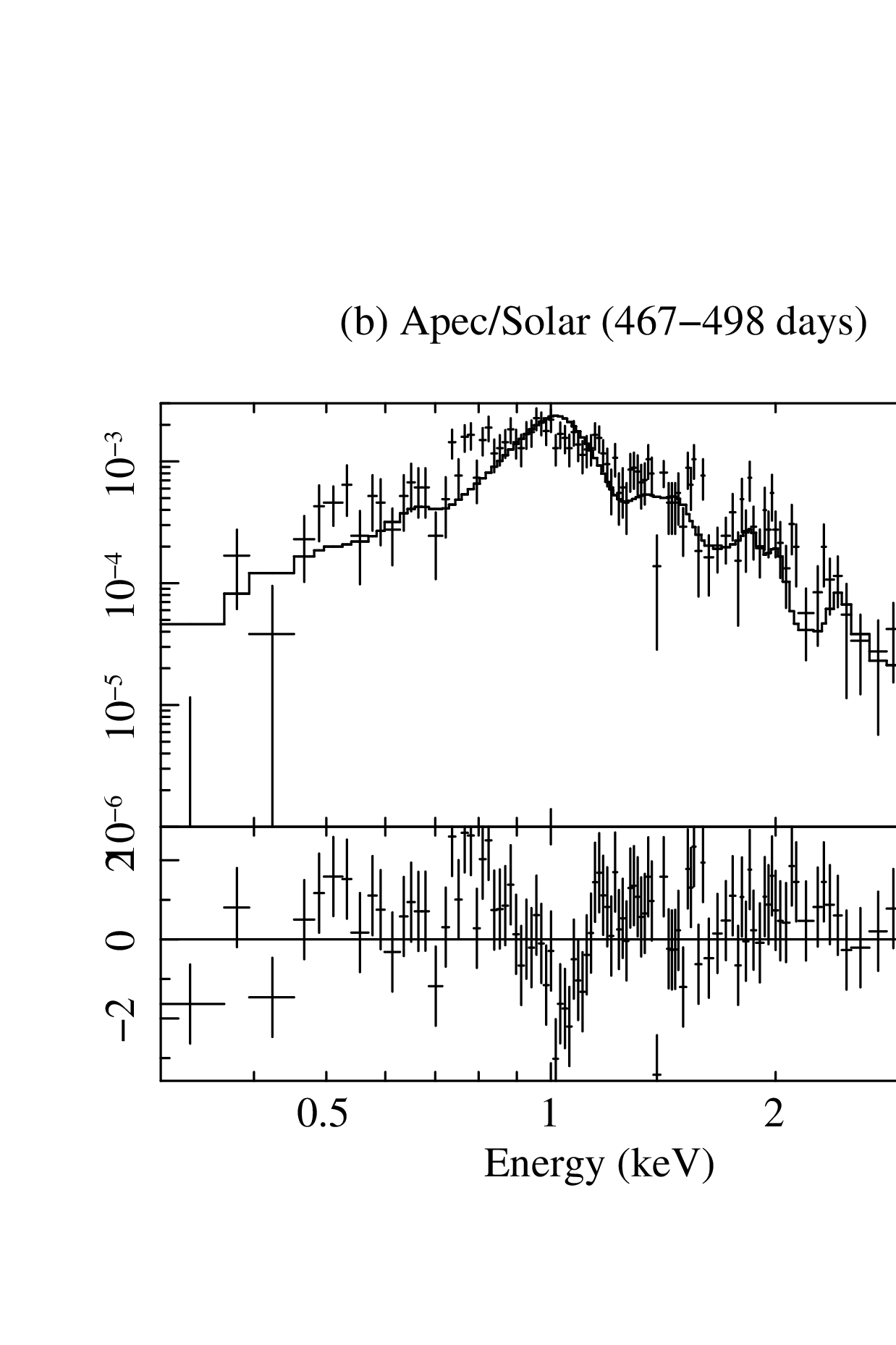}
        \end{minipage}\\
\hspace{-1.5cm}
\vspace{-3cm}
        \begin{minipage}[]{0.45\textwidth}
                \epsscale{0.9}
                \plotone{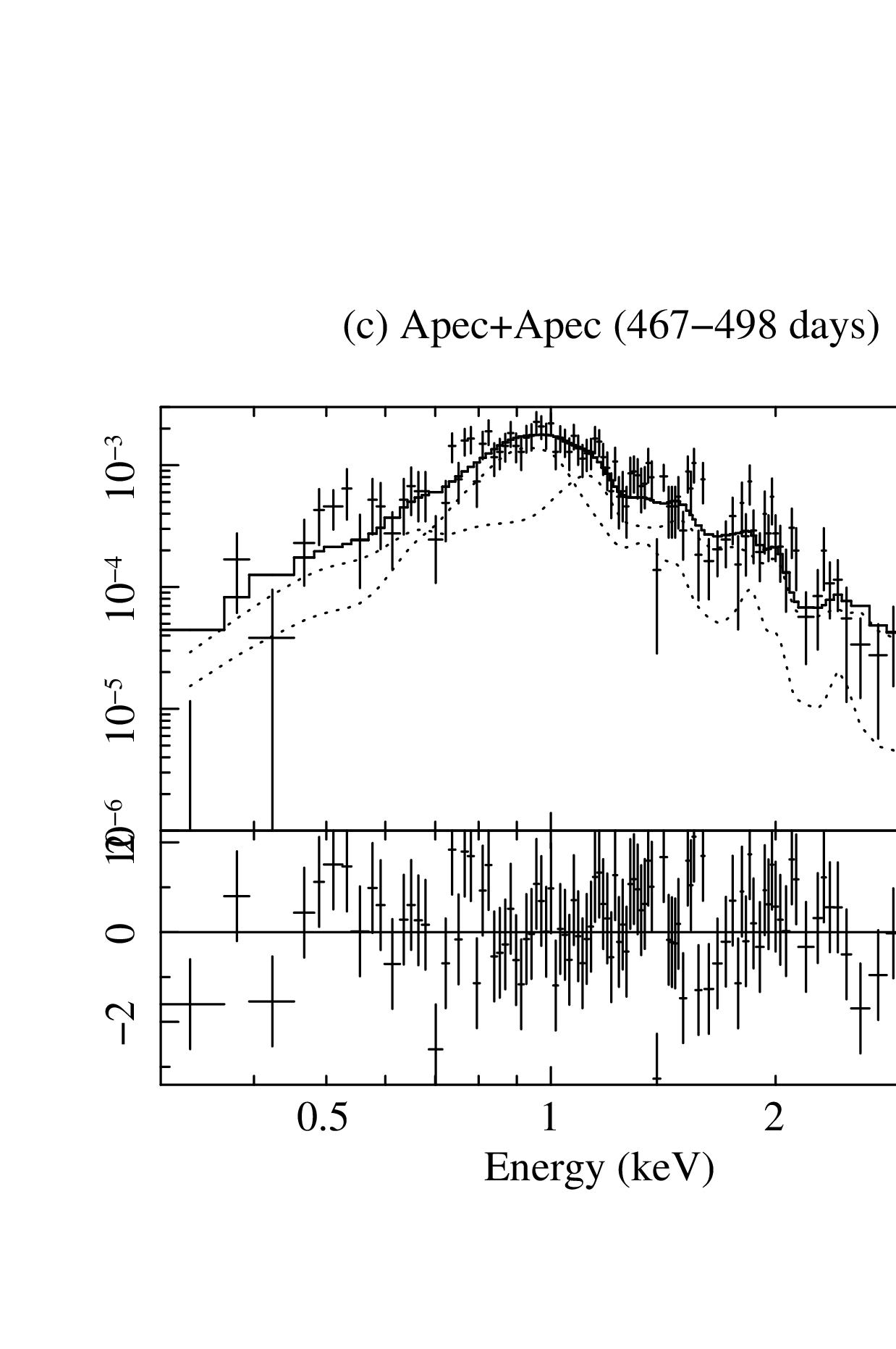}
        \end{minipage}
\end{center}
\vspace{1.5cm}
\caption
{Model fits to the {\em Chandra} spectrum integrated between 467 - 498 days, using (a) a power law model, (b) a one-component, H-rich/solar metallicity {\bf apec} model and (c) a two-component, H-rich/solar metallicity {\bf apec}+{\bf apec} model. For (c), the contributions from the low- and high-temperature components are shown by the dotted curves. 
\label{fig3}}
\end{figure}

\begin{deluxetable*}{cccccc}
 \tabletypesize{\scriptsize}
 \tablecaption{Spectral Model Fits to SN 2011dh at $\sim 500$ days. \tablenotemark{a}
 \label{tab1}}
 \tablewidth{0pt}
 \tablehead{
   \colhead{Model}
 & \colhead{$\chi_{\rm red}^2$/d.o.f. (n.h.p.)}
 & \colhead{$N_{\rm H}$ ($10^{20}$ cm$^{-2}$)}
 & \colhead{Parameter ($\Gamma/kT$ (keV))}
 & \colhead{Parameter (abundance)}
 & \colhead{L$_{\rm X}$ ($10^{37}$ erg s$^{-1}$)}
}
\startdata
PL & $1.23/228 \ (9.5 \times 10^{-3})$ & $47 \pm 6.5$ & $4.59 \pm 0.35$ & \nodata & 93.4\\
apec & $1.17/229 \ (3.9 \times 10^{-2})$ & $1.8$ (fixed) & $1.26 \pm 0.03$& 1.0 (fixed)& 4.5\\
apec & $1.03/227 \ (0.37)$ & $6.8 \pm 2.9$ & $1.13 \pm 0.08 $ & $0.22 \pm 0.07$ & 6.1\\
apec & $1.05/228 \ (0.30)$ & $1.8$ (fixed) & $1.20 \pm 0.05$ & $0.30 \pm 0.07$& 5.3\\
vapec & $1.02/227 \ (0.42)$ & $9.9 \pm 2.9$ & $1.03 \pm 0.04$ & $17.4 \pm 4.2$ (He) & 6.6\\
vapec & $1.04/228 \ (0.31)$ & $1.8$ (fixed) & $1.20 \pm 0.05$ & $9.9 \pm 2.9$ (He)& 5.3\\
apec & $0.95/226\ (0.71)$ & $3.1 \pm 2.8$ & $0.96 \pm 0.06$ & $1.0$ (fixed) & 2.5\tablenotemark{b}\\
+apec & \nodata & \nodata & $2.4 \pm 0.5$ & \nodata & 3.3\tablenotemark{b}\\
apec & $0.94/227\ (0.72)$ & $1.8$ (fixed) & $0.97 \pm 0.05$ & $1.0$ (fixed) & 2.3\tablenotemark{c}\\
+apec & \nodata & \nodata & $2.5 \pm 0.5$ & \nodata & 3.3\tablenotemark{c}\\
apec & $0.94/226\ (0.73)$ & $1.8$ (fixed) & $0.96 \pm 0.05$ & $1.82 \pm 0.95$ & 2.3\tablenotemark{d}\\
+apec & \nodata & \nodata & $2.8 \pm 0.7$ & \nodata & 3.5\tablenotemark{d}\\
vapec & $0.94/227\ (0.73)$ & $1.8$ (fixed) & $0.97 \pm 0.05$ & $10.0$ (He,fixed) & 2.3\tablenotemark{e}\\
+vapec & \nodata & \nodata & $2.5 \pm 0.5$ & $+4.0$ (others, fixed) & 3.3\tablenotemark{e}\\
\enddata
\tablenotetext{a}{The error is shown for $1\sigma$.}
\tablenotetext{b}{Total luminosity is $5.9 \times 10^{37}$ erg s$^{-1}$.}
\tablenotetext{c}{Total luminosity is $5.7 \times 10^{37}$ erg s$^{-1}$.}
\tablenotetext{c}{Total luminosity is $5.8 \times 10^{37}$ erg s$^{-1}$.}
\tablenotetext{e}{Total luminosity is $5.7 \times 10^{37}$ erg s$^{-1}$.}
\end{deluxetable*}

{\bf Power law Model (PL): } It does not give an acceptable fit (null hypothesis probability, n.h.p., of $\sim 10^{-2}$; Fig. 3a). It further requires large $N_{\rm H}$, $\sim 5 \times 10^{21}$ cm$^{-2}$, more than one order of magnitude larger than that estimated for the early phase spectrum \citep{soderberg2012}. Given that a non-thermal emission is expected to originate in the forward shock \citep{chevalier2003}, it is quite unlikely that $N_{\rm H}$ {\em increases} with time (see also below). Furthermore, the spectrum is very soft, requiring the spectral index of $\Gamma \sim 4.5$. This is incompatible to the non-thermal mechanism, either the IC or synchrotron emission. In sum, we find that the PL model is not appropriate for the spectrum at $\sim 500$ days.

{\bf Thermal plasma model (apec): }  We next try the {\bf apec} model, i.e., optically thin thermal emission in collisional ionization equilibrium \citep{smith2001}\footnote{While we focus on the {\bf apec} model in the following analyses, we have repeated the same analyses using the {\bf mekal} model, confirming that similar results are obtained.}. If we fix the abundance to the solar values, the column density is consistent with the Galactic value. With the solar abundance, the fit is statistically poor (n.h.p $\sim 3.9 \times 10^{-2}$; Fig. 3b), but it indicates that the peaks seen in the spectrum (e.g., $\sim 1.8$ keV) are real, arising from metal emission lines (compare Fig. 3a and Fig. 3b). 

If we take the abundance as a free parameter (while fixing the relative abundance to the solar), a statistically acceptable fit is obtained for the sub-solar abundance ($Z = 0.22 \pm 0.07 Z_{\odot}$ for $N_{\rm H}$ taken as a free parameter, or $Z = 0.30 \pm 0.08 Z_{\odot}$ for $N_{\rm H}$ fixed to the Galactic value\footnote{The F-test probability of varying $N_{\rm H}$ is 0.037, and thus the model with $Z = 0.22 Z_{\odot}$ is just marginally better than the one with $Z = 0.30 Z_{\odot}$}). On the other hand, the metallicity at the progenitor vicinity has been inferred to be of sub-solar to solar, with nearby HII region metallicities spanning $Z \sim 0.6 Z_{\odot}$ to $\sim Z_{\odot}$ \citep{bresolin2004,vandyk2011,sahu2013}. With no direct indication of such a low metallicity in the progenitor, lower than the surrounding environment, we regard this interpretation quite unlikely. 

\begin{figure}
\begin{center}
\vspace{-2.0cm}
\hspace{-1.5cm}
        \begin{minipage}[]{0.45\textwidth}
                \epsscale{0.9}
                \plotone{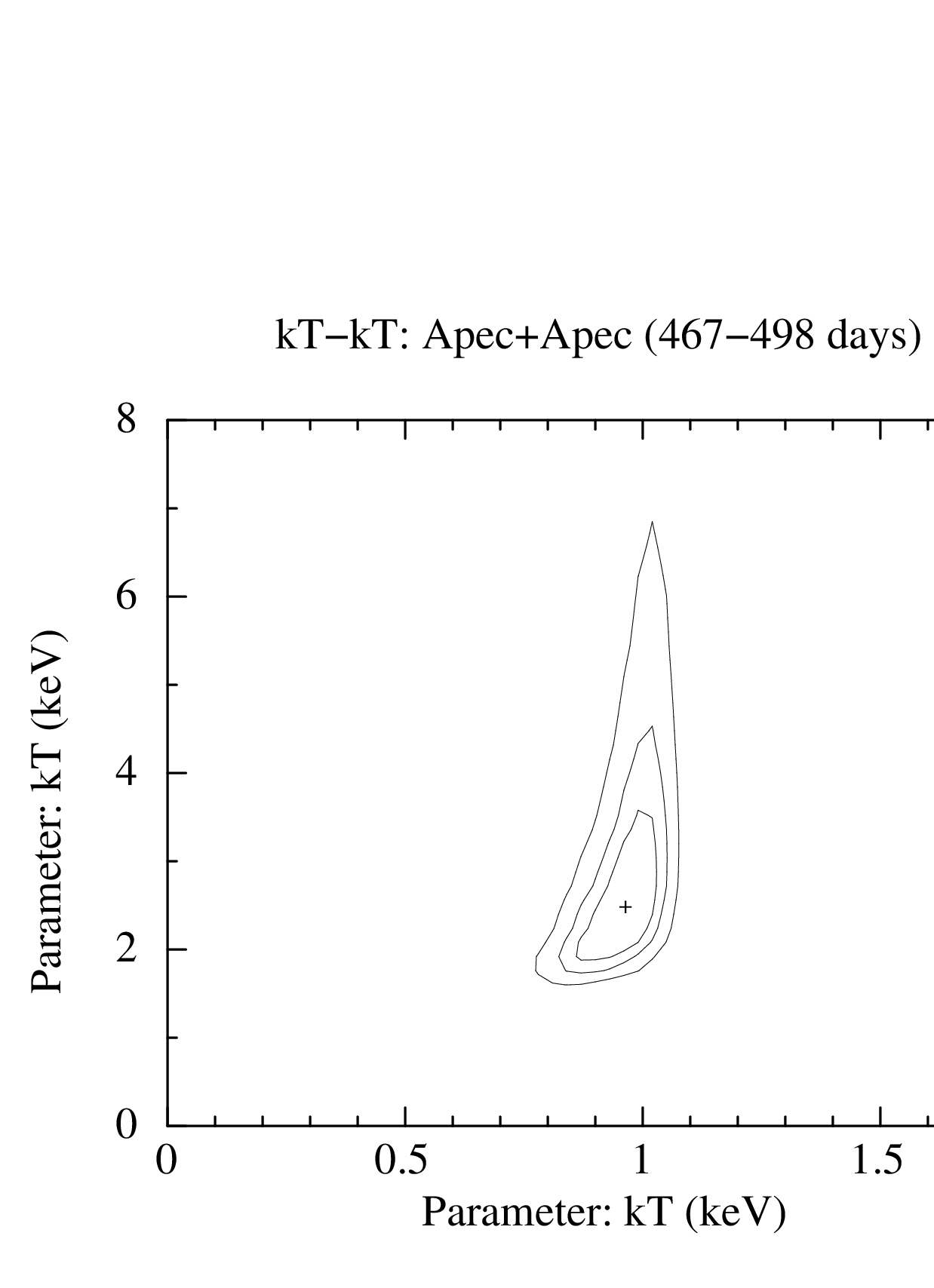}
        \end{minipage}
\end{center}
\caption
{Confidence contour of kT vs. kT in the two-component H-rich model. The contours are shown for the confidence levels of 68\%, 90\%, and 99\%. 
\label{fig4}}
\end{figure}

\begin{figure*}
\begin{center}
\vspace{-2cm}
\hspace{-0.7cm}
        \begin{minipage}[]{0.45\textwidth}
                \epsscale{0.8}
                \plotone{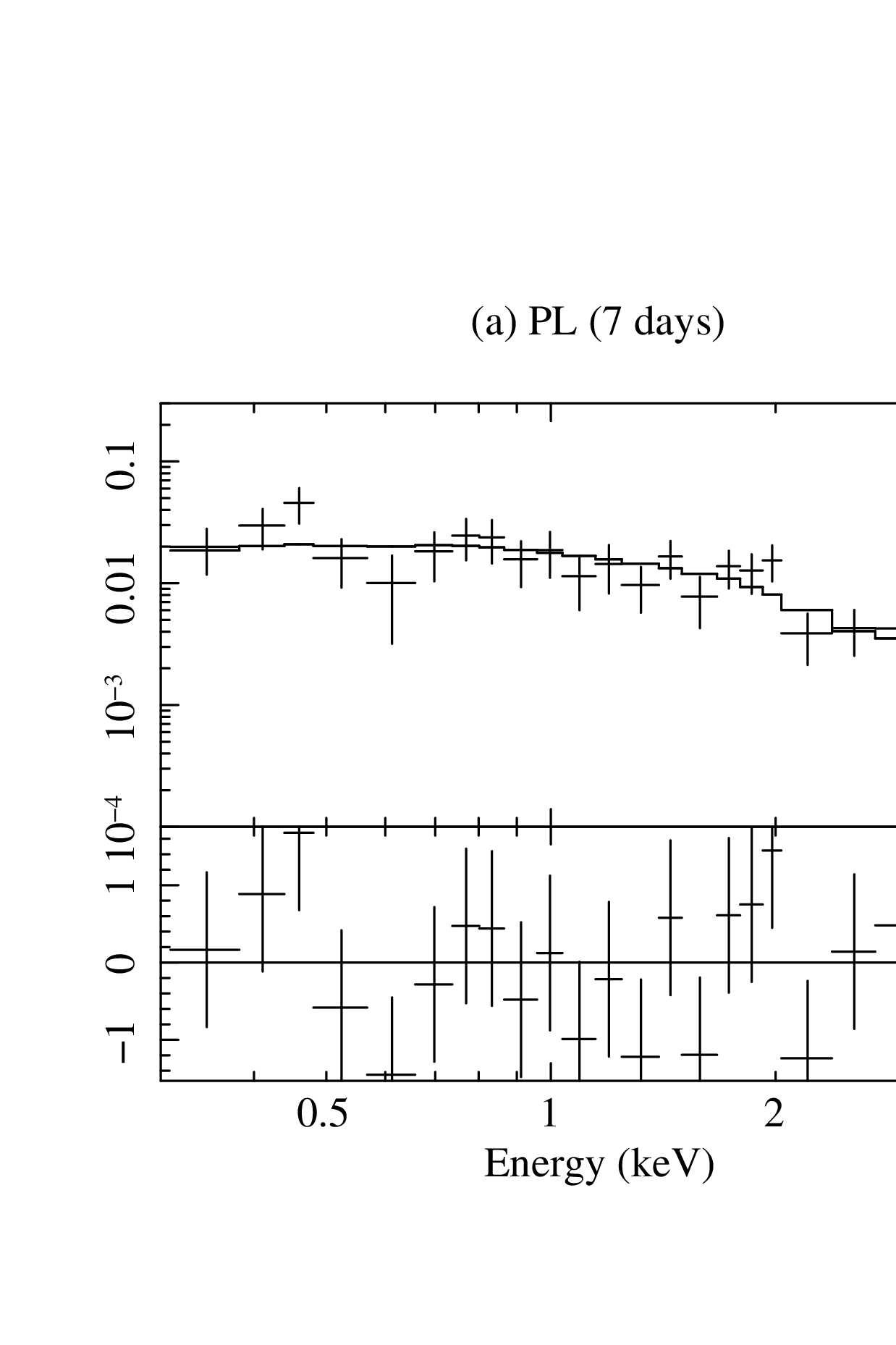}
        \end{minipage}
\hspace{+0.7cm}
        \begin{minipage}[]{0.45\textwidth}
                \epsscale{0.8}
                \plotone{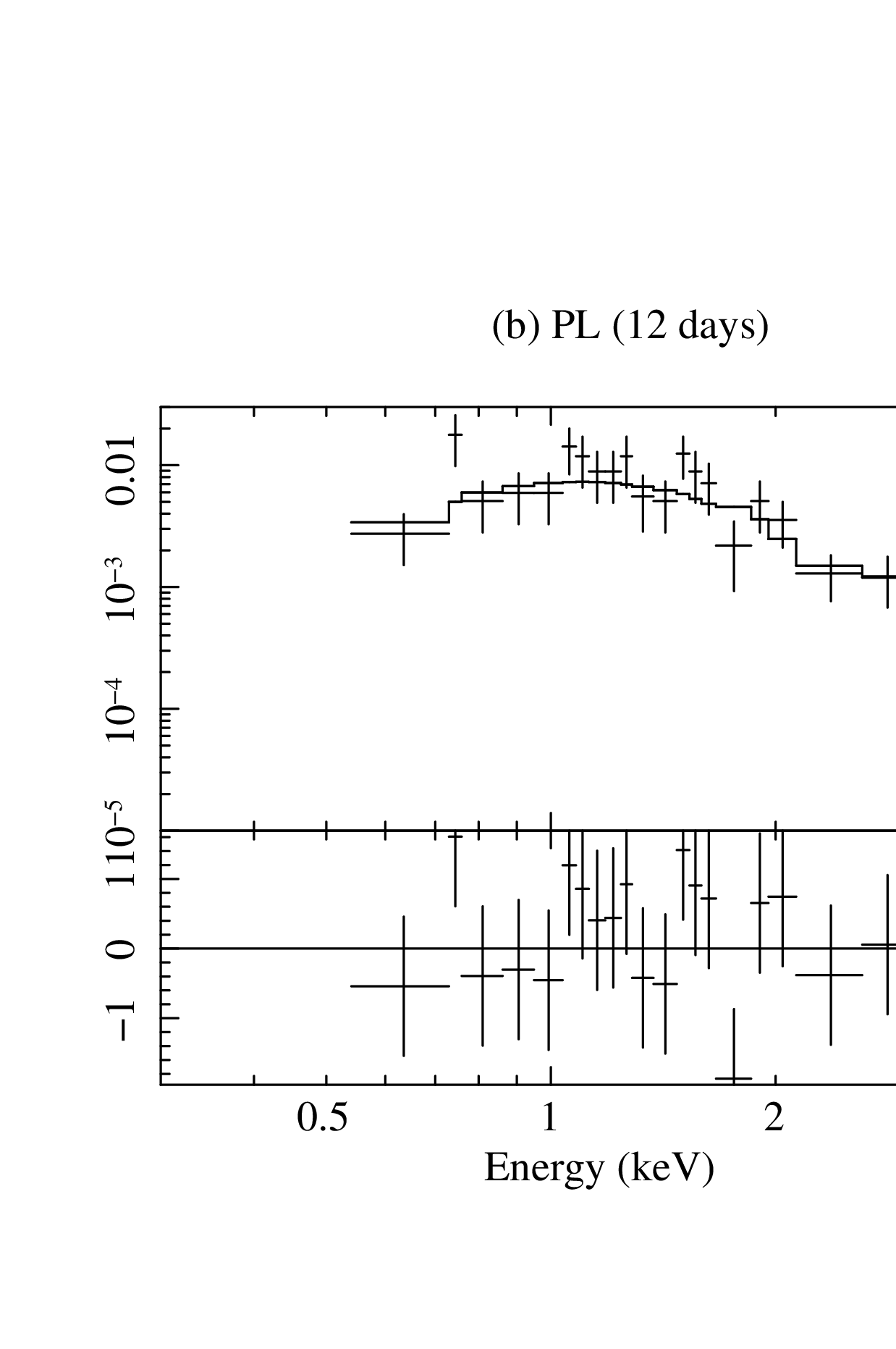}
        \end{minipage}\\
\vspace{-2cm}
\hspace{-0.7cm}
        \begin{minipage}[]{0.45\textwidth}
                \epsscale{0.8}
                \plotone{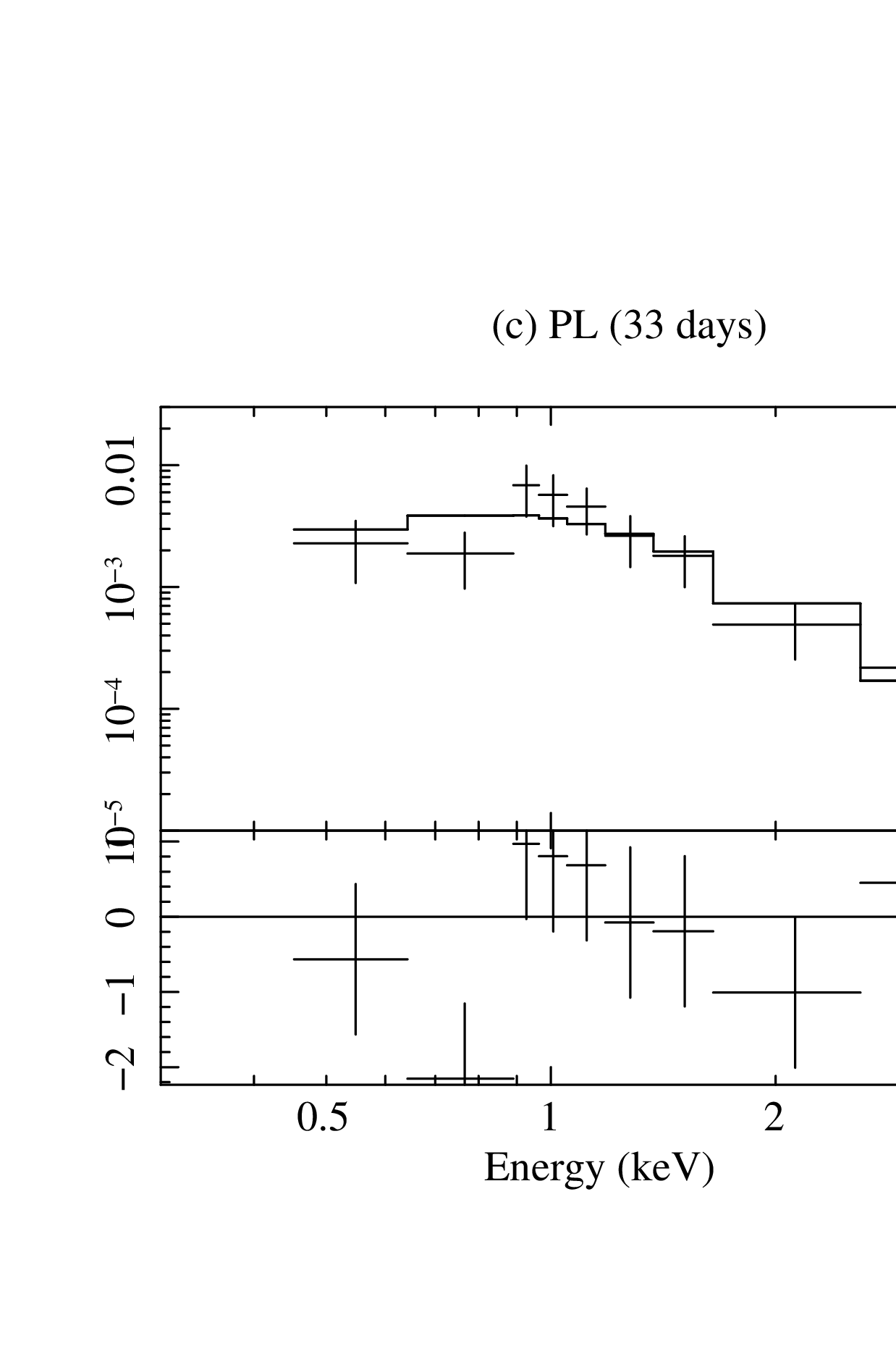}
        \end{minipage}
\hspace{+0.7cm}
        \begin{minipage}[]{0.45\textwidth}
                \epsscale{0.8}
                \plotone{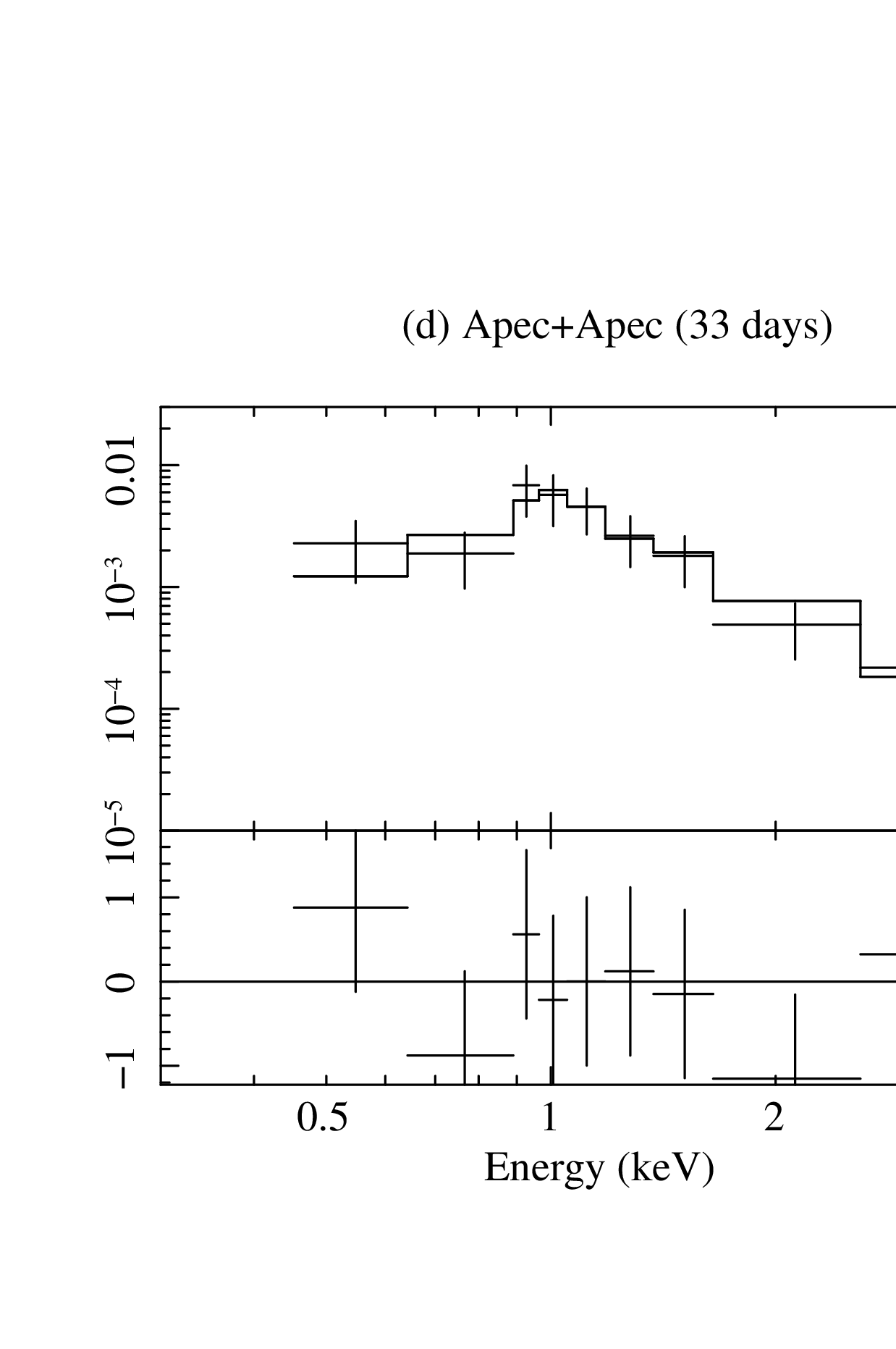}
        \end{minipage}
\end{center}
\vspace{-1cm}
\caption
{Early-phase spectra. Shown here are (a) the {\em XMM-Newton} spectrum at 7 days with a power law model, (b) the {\em Chandra} spectrum at 12 days with a power law model, one at 33 days (c) with a power law model or (d) with a two-component H-rich {\bf apec+apec} model. 
\label{fig5}}
\end{figure*}

{\bf He-rich thermal plasma model (vapec): } In the thermal plasma model using {\bf vapec}, we also find that the He-rich model provides similar results, where the He abundance is varied while the abundances of other elements relative to H are fixed to the solar values. An acceptable fit is obtained for the He abundance of $17.4 \pm 4.2$ (for $N_{\rm H}$ taken as a free parameter) or $9.9 \pm 2.9$ (for the Galactic column density)\footnote{Thawing $N_{\rm H}$ in the fit results in the F-test probability of 0.04.}. We however note that while a large abundance of He in the thermal-plasma model fit, using {\bf Xspec}, is sometimes regarded to be a {\em direct} signature of the emission from the He-rich layer \citep[e.g., ][]{smith2007}, one has to be careful to interpret such a result. This model does require the sub-solar metallicity like the H-rich envelope model. With He and other  abundances {\em relative to H} being 10 solar and 1 solar, this corresponds to the metallicity of $\sim 0.24Z_{\odot}$ - this is indeed very similar to the H-rich, sub-solar model as described above. Namely, the thermal emission from the He-rich thermal plasma is {\em just consistent} with the data, but such a model is not in particular favored/disfavored than the H-rich model. Following our argument against the sub-solar metallicity, we regard this interpretation unlikely.

{\bf Two-component thermal plasma model (apec+apec): } We then investigate if there is an acceptable model with solar metallicity. In doing this, we introduce an absorbed two-component thermal plasma model ({\bf apec+apec}) (Fig. 3c). This model, with the solar abundance, provides an acceptable fit. The model with $N_{\rm H}$ and $kT$ taken as free parameters (while fixing the abundances to the solar values) results in n.h.p. of $\sim 0.7$ and the F-test probability of $\sim 1.6 \times 10^{-11}$ as compared to the corresponding one-component solar-metallicity {\bf apec} model. The column density is consistent with the Galactic value. Varying the abundance (while tying those of high- and low-temperature components) results in $Z = 1.82 \pm 0.95 Z_{\odot}$ and the F-test probability of $\sim 1$, thus the two-component model does not favor the sub-solar abundance -- therefore hereafter our analysis is based on the solar-abundance model. A He-rich two-component model ({\bf vapec+vapec}) provides very similar results to the (H-rich) solar abundance model. 

In the two-component model, the high- and low-temperature components have the temperature of $kT \sim 2.5$ and $\sim 1$keV, respectively, as shown in the confidence contours (Fig. 4). The contributions of individual components to the total (unabsorbed) luminosity are comparable, with the high-temperature component larger by $\sim 30$\%. As we show below, the temperature of the high temperature component is much lower than expected for the forward shock, thus we attribute both of the two-components to the reverse shock -- Namely a situation where the emission is mostly originated from the reverse shock, but with variation in the temperature behind the reverse shock \citep[see, e.g., ][]{nymark2009}. Such variation in the temperature within the shocked ejecta can, for example, naturally be realized as follows: (1) The outer part is hit by the reverse shock earlier, and thus suffers from some cooling already. This may correspond to the low-temperature component. (2) The ejecta may have fluctuations in the density, perhaps with clumps, and then the high density region (or clumps) will result in a lower reverse shock velocity and lower temperature. Irrespective of the physical interpretation, we regard this two-component (solar-metallicity) model as our best-fit model.

We apply the best-fit H-rich two-component model to individual spectra during 467--498 days. Here, we fix the temperature and the abundances to the values obtained through the fit to the integrated spectrum, but vary only the normalization. The relative normalization between the high- and low-temperature components is fixed. Due to a low photon statistic we perform the C-statistics \citep{cash1979} rather than $\chi^2$ test. We thereby found that this model is statistically acceptable for all the individual spectra\footnote{We confirmed that this is not sensitive to a particular model adopted in the analysis, by doing the same analysis for the two-component He-rich model and for the one-component sub-solar H-rich/He-rich models.}.

For the cross check of our analysis methods, we also reanalyze the early-phase data obtained by {\em XMM-Newton} \citep{campana2012,sasaki2012} and {\em Chandra} \citep{soderberg2012}, as shown in Figure 5. Given the low photon statistics in these data (especially the {\em Chandra} data), the best model cannot be selected purely on statistic basis. Following the previous study, we first perform an absorbed power law fit. Within the power law fit, we do not find significant absorption in addition to the Galactic value. For the spectra at 7, 12 and 33 days since the explosion, the best-fit power law models give $\Gamma = 1.2 \pm 0.1$, $1.5 \pm 0.2$ and $2.3 \pm 0.3$, respectively (assuming only the Galactic absorption). These are consistent with the previous study \citep{campana2012,sasaki2012,soderberg2012}, showing a spectral softening with time which may be understood as a combination of a decaying harder component and a persistent softer component \citep[][see \S 4 for more details]{sasaki2012}. 

We also try the thermal plasma model for the early-phase data. If we try the one-component {\bf apec} model with the abundance as a free parameter, we obtain $kT = 1.5 \pm 0.4$ keV and $Z < 0.45 Z_{\odot}$ ($1\sigma$) for the data at 33 days -- it prefers the low metallicity within the one-component prescription, as is similar to the fit to the data at 500 days, suggesting that the X-ray emission at 33 days could actually be explained by the same mechanism as 500 days. The same one-component model for the spectrum at 12 days results in $kT = 6.1 \pm 1.7$ keV, quite different from those at days 33 and 500. The {\em XMM-Newton} spectrum at 7 days is even harder and incompatible to the thermal emission from the plasma with $kT \lsim 10$ keV. In sum, we conclude that the X-ray emission at 33 days could be largely explained by the same X-ray emission mechanism with that at 500 days, while the earlier data at 7 and 12 days would require a different mechanism. The two-component model, with a larger uncertainty than the one component model due to larger degree of freedom for the poor photon count data, leads to similar conclusions (e.g., $kT = 1.2 \pm 0.4$ keV for the low temperature component at 33 days). This point will be discussed further in \S 4. 

\section{Constraints on Properties of the Ejecta and CSM}

The X-ray light curve of SN 2011dh (assuming the distance of 8.4 Mpc) is shown in Figure 6. We extract an unabsorbed flux of SN 2011dh in $0.3-8$ keV, averaged in 467--498 days since the explosion, based on our best fit model to the integrated spectrum -- Namely, the two-component H-rich, solar-abundance thermal plasma model with only the Galactic column density. The derived X-ray luminosity is not sensitive to the particular model used here (see Tab. 1). 
The fluxes in individual epochs were also extracted by applying the same model (see \S 2). For the individual points, the errors account only for the global normalization (e.g., omitting that comes from the relative normalization between the two-components), therefore the errors are rather optimistic; therefore we do not claim significant evolution between the short period in 467 - 498 days. The fluxes extracted with the power law models for the early-phase data are consistent with the previous study \citep{campana2012,sasaki2012,soderberg2012}. 

\begin{figure}
\begin{center}
        \begin{minipage}[]{0.45\textwidth}
                \epsscale{1.2}
                \plotone{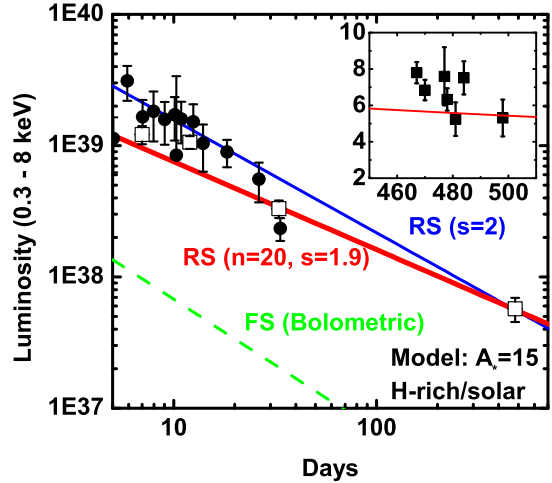}
        \end{minipage}
\end{center}
\vspace{-0.5cm}
\caption
{Unabsorbed X-ray light curve of SN IIb 2011dh. The {\em Chandra} and {\em XMM-Newton} points are shown by open squares. We also plot the published data from \citet{soderberg2012} by filled circles. The inset shows the light curve in the late-phase extracted from individual spectra (filled squares). For the individual points, the errors account only for the global normalization (e.g., omitting that comes from the relative normalization between the two-components), therefore the errors are rather optimistic; therefore we do not claim significant evolution between the short period in 467 - 498 days. Unabsorbed X-ray light curves are shown for the reverse shock with $A_{*} = 15$, with $(n, s) = (20, 1.9)$ (red/thick solid) and $(20, 2.0)$ (blue/thin solid), where $n$ and $s$ are defined by $\rho_{\rm ej} \propto v^{-n}$ and $\rho_{\rm CSM} \propto r^{-s}$. The forward shock contribution is also shown (green dotted). 
\label{fig6}}
\end{figure}

The forward shock velocity ($V_{\rm FS}$) has been constrained by {\em VLBI} observations. The average speed up to $179$ days is $21,000 \pm 7000$ km s$^{-1}$@\citep{bietenholz2012}. Thus the forward shock velocity at 500 days should be within the range of $18,500 - 21,000$ km s$^{-1}$ ($\pm 7,000$ km s$^{-1}$) taking into account the deceleration; the lower value is obtained for a canonical model for the deceleration \citep{chevalier1982}, $V_{\rm FS} \propto t^{-0.12}$ (for $n \sim 10$ and $s \sim 2$, where $n$ is the power law index of the ejecta outer density distribution given as $\rho_{\rm ej} \propto v^{-n}$, and $s$ is for the CSM density distribution given as $\rho_{\rm CSM} \propto r^{-s}$) while the larger value assumes free expansion \citep[for either the large value of $n$ or very low CSM density; ][]{chevalier1982,maeda2013b}. The electron temperature behind the forward shock is given as follows: 
\begin{eqnarray}
T_{\rm FS} & = & 2.27 \times 10^{9} \mu_{s} \left(\frac{V_{\rm FS}}{10,000 \ {\rm km s}^{-1}}\right)^2 \ {\rm K} \nonumber \\
& \sim & \left\{
\begin{array}{rl}
(3.6 - 4.7) \times 10^{9} \ {\rm K,} & \ \mbox{for $\mu_{s} = 0.61$} \ ,\\
(7.8 - 10.0) \times 10^{9} \ {\rm K,} & \ \mbox{for $\mu_{s} = 1$} \ .
\end{array}\right.
\end{eqnarray}
Here the mean mass per particle (in a.m.u.) is described by $\mu_{s}$: $\mu_{s} = 0.61$ for the solar metallicity (i.e., H-rich composition), while $\mu_{s} \sim 1$ for the He-rich composition [$n(H) = n(He) = 1$]. This estimate assumes the equipartition between thermal ions and electrons, and thus could be an overestimate of the electron temperature behind the (low-density) forward shock by a factor of $\sim 2$ \citep{fransson1996}. In any case, the  electron temperature behind the forward shock  is clearly too high to explain the observed X-ray emission at 500 days. 

The only remaining possibility for the relatively low temperature ($\sim 1 - 2.5$ keV) thermal emission is the one from the reverse shock. The electron temperature behind the reverse shock can be estimated as follows \citep[e.g., ][]{chevalier2003}: 
\begin{equation}
T_{\rm RS} = \frac{(3-s)^2}{(n-3)^2} T_{\rm FS} \ , 
\end{equation}
under the assumption of the electron-ion equipartition (which will be justified later). Hereafter we adopt the two component model(s) in our analysis, while adopting the one-component sub-solar model does not change our conclusions (\S 5). For $s \sim 2$ expected for the steady mass loss as was also indicated by the radio observations \citep{soderberg2012,maeda2012}, we find that either $n \sim 20$ (H-rich case) or $n \sim 25$ (He-rich case)  is necessary so that the temperature behind the reverse shock explains the characteristic energy in the observed X-ray emission (Fig. 7). This is quite steep as compared to a fiducial value ($n \sim 10$) frequently assumed for SE-SNe with a compact progenitor \citep[e.g., ][]{chevalier2006}. Indeed there is evidence that SN IIb 1993J, from a giant progenitor with an extended envelope, had the steep density gradient at its outermost layer ($n \gsim 20$), inferred from the hydrodynamic explosion model \citep{suzuki1995} and X-ray analysis \citep{fransson1996}. Therefore, it is likely that SN IIb 2011dh from the YSG progenitor shares a property in the outermost layer similar to SN IIb 1993J rather than other SE-SNe (SNe Ib/c and a part of SNe IIb). 

\begin{figure}
\begin{center}
        \begin{minipage}[]{0.45\textwidth}
                \epsscale{1.2}
                \plotone{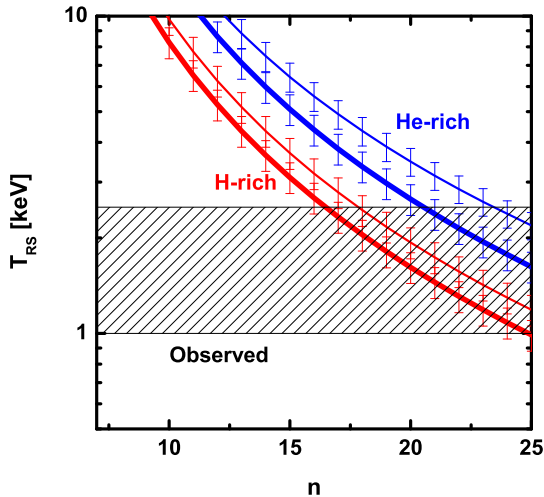}
        \end{minipage}
\end{center}
\caption
{A constraint on the ejecta property. The temperature behind the reverse shock as a function of $n$ (the power law index of the outer density distribution) is shown. The constraint is shown for H-rich composition (red) and He-rich composition (blue). In each case, a difference value for $s$ (the power law index of the CSM distribution) is shown, for $s = 2.0$ (thick) and $s=1.9$ (thin), covering the range of uncertainty in the mass loss history (see \S 4 for details). The errors are for the VLBI measurement of the forward shock velocity at 179 days ($\pm 7,000$ km s$^{-1}$).
\label{fig7}}
\end{figure}

The X-ray luminosity is roughly described by the following expression: $L \sim \Lambda n_{\rm e} n_{\rm H} V$, where $\Lambda$ is the cooling function (erg s$^{-1}$ cm$^{3}$), $n_{\rm e}$ is the electron density (cm$^{-3}$), $n_{\rm H}$ is the hydrogen density (cm$^{-3}$), $V$ is the volume (cm$^{3}$) of the emitting region. Since the cooling function, taking into account the metal content, is given in plasma models in XSPEC, we can measure $n_{\rm e} n_{\rm H} V$ (as a norm) by the spectral fitting. For the high temperature component in the two-component model(s), we find the following: 
\begin{equation}
n_{\rm e, h} n_{\rm H, h} V_{\rm h} 
\sim \left\{
\begin{array}{rl}
2.4\times 10^{60} \ {\rm cm}^{-3} \ , & \ \mbox{for H-rich composition} \ ,\\
6.9\times 10^{59} \ {\rm cm}^{-3} \ , & \ \mbox{for He-rich composition} \ .
\end{array}\right.
\end{equation}
For the low temperature component, we find the following: 
\begin{equation}
n_{\rm e, l} n_{\rm H, l} V_{\rm l} 
\sim \left\{
\begin{array}{rl}
1.2\times 10^{60} \ {\rm cm}^{-3} \ , & \ \mbox{for H-rich composition} \ , \\
2.9\times 10^{59} \ {\rm cm}^{-3} \ , & \ \mbox{for He-rich composition} \ .
\end{array}\right.
\end{equation}
Here the subscripts h and l refer the high- and low-temperature components, respectively. 
Taking into account the different compositions, we can rewrite these expressions in terms of the density behind the reverse shock ($n_{\rm RS}$) and the mass swept up by the reverse shock ($M_{\rm RS}$). Doing this requires another constraint -- for simplicity first we assume that the densities behind the reverse shock are (roughly) the same for the high-temperature and low-temperature components, which would correspond to the situation that the high-temperature component reflects a component freshly injected into the reverse shock while the low temperature component reflects the ejecta which was hit by the reverse shock earlier and experienced some cooling already. We then obtain the following constraint: 
\begin{equation}
n_{\rm RS} \frac{M_{\rm RS}}{M_{\odot}}  
\sim \left\{
\begin{array}{rl}
3000 \ {\rm cm}^{-3} \ , & \ \mbox{for H-rich composition} \ , \\
2700 \ {\rm cm}^{-3} \ , & \ \mbox{for He-rich composition} \ .
\end{array}\right.
\end{equation}
Here, the subscript RS is for the quantities behind the reverse shock. 

With $R \sim 3 \times 10^{16}$ cm at 179 day \citep{bietenholz2012}, the radius reached by the ejecta at 500 days is estimated to be $R \sim 8 \times 10^{16}$ cm. This is for $n=20$, but it is anyway not sensitive to the value of $n$. With $R \sim 8 \times 10^{16}$ cm, the mass swept up by the reverse shock is (assuming $s = 2$): 
\begin{equation} 
M_{\rm RS} \sim 3.5 M_{\rm FS} \sim 8.8 \times 10^{-4} A_{*} M_{\odot} \ .
\end{equation} 
Here the coefficient in the conversion is from \citet{chevalier1982}. The density behind the reverse shock, with $R \sim 8 \times 10^{16}$ cm, is estimated as follows: 
\begin{eqnarray}
n_{\rm RS} (R) & = & 4 \frac{(n-3)(n-4)}{(3-s)(4-s)} n_{\rm CSM} (R) \nonumber \\
& \sim & \left\{
\begin{array}{rl}
2.5 \times 10^{4} A_{*} \ {\rm cm}^{-3} \ , & \ \mbox{for H-rich composition} \ , \\
1.0 \times 10^{4} A_{*} \ {\rm cm}^{-3} \ , & \ \mbox{for He-rich composition} \ .
\end{array}\right.
\end{eqnarray}
Here we assume $n = 20$ and $s=2$. 

\begin{figure*}
\begin{center}
        \begin{minipage}[]{0.3\textwidth}
                \epsscale{1.2}
                \plotone{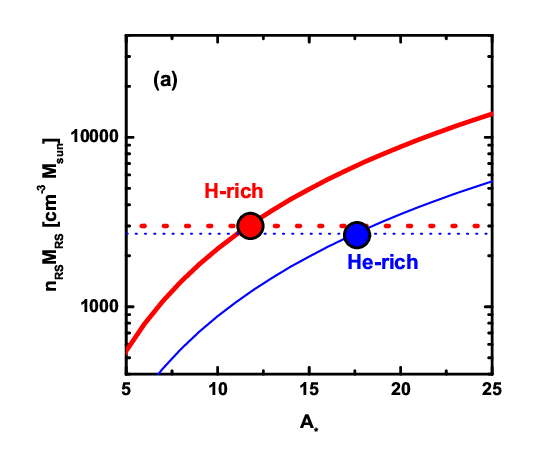}
        \end{minipage}
        \begin{minipage}[]{0.3\textwidth}
                \epsscale{1.2}
                \plotone{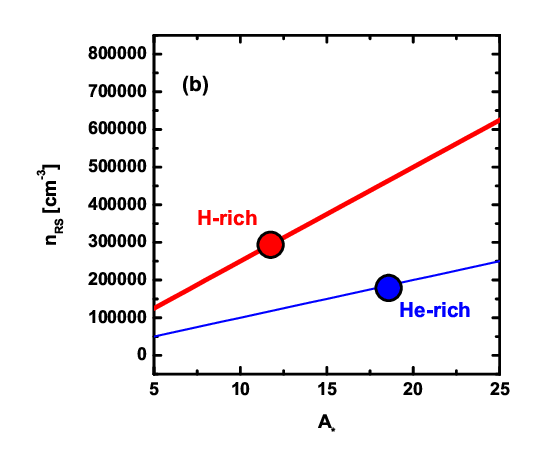}
        \end{minipage}
        \begin{minipage}[]{0.3\textwidth}
                \epsscale{1.2}
                \plotone{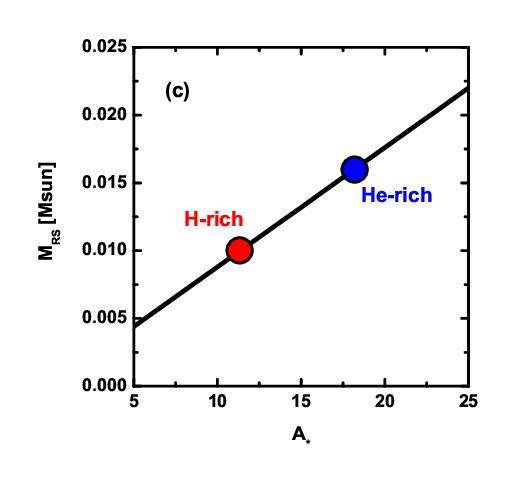}
        \end{minipage}
\end{center}
\caption
{A constraint on the CSM property. (a) A constraint on the CSM density ($A_{*}$) through the product $n_{\rm RS} M_{\rm RS}$. The solid curves show the expected behaviors of $n_{\rm RS} M_{\rm RS}$ as a function of $A_{*}$ (red for the H-rich model and blue for the He-rich model), while the horizontal dotted lines show the observationally derived constraints (for the H-rich/He-rich models). (b) The density behind the reverse shock as a function of $A_{*}$, where the points are our solutions from the constraint by $n_{\rm RS} M_{\rm RS}$. (c) The mass of the swept up ejecta as a function of $A_{*}$. 
\label{fig8}}
\end{figure*}

By combining these expressions (5--7), we derive $A_{*} \sim 12$ for the H-rich composition and $A_{*} \sim 18$ for the He-rich composition (Fig. 8). Namely, the result, $A_{*} \sim 15$, is not sensitive to the unknown composition within the outer envelope where the reverse shock is penetrating. The mass swept up by the reverse shock at $\sim 500$ days is $M_{\rm RS} \sim 0.013 M_{\odot}$, again insensitive to the composition. The mass loss rate is thus $\dot M \sim 3 \times 10^{-6} M_{\odot}$ yr$^{-1}$ $(v_{w}/20 \ {\rm km} \  {\rm s}^{-1})$, corresponding roughly the one at $\sim 1,300 \times (v_{w}/20 \ {\rm km} \ {\rm s}^{-1})^{-1}$ yrs before the explosion\footnote{Hereafter we normalize the mass loss rate by the wind velocity of $v_{w} \sim 20$ km s$^{-1}$, as is the same with the one adopted for an YSG progenitor by \citet{krauss2012}}. 

To estimate an uncertainty in the above derivation, we also consider a case where $n_{\rm RS, h} \ne n_{\rm RS, l}$. A physically motivated case is that the difference reflects a pre-shocked density variation, and in this situation we can relate the densities and temperatures behind the reverse shock by $T_{\rm RS, l}/T_{\rm RS, h} \propto (n_{\rm RS, l}/n_{\rm RS, h})^{-1}$. From this, we have a constraint as $n_{\rm RS, h} \sim 0.4 n_{\rm RS, l}$. In this situation $M_{\rm RS, h} \sim 5 M_{\rm RS, l}$ and thus the high-temperature component dominates the mass in the emitting region. In this case, following the same procedure for the case with $n_{\rm RS, h} = n_{\rm RS, l}$, we estimate $A_{*} \sim 10$ for $n_{\rm RS, h} \ne n_{\rm RS, l}$, not far from the estimate above. Since the two situations mentioned above cover the opposite situations in the possible difference between $n_{\rm RS, l}$ and $n_{\rm RS, h}$, we conclude that our derivation of $A_{*} \sim 15$ is robust.

This solution ($A_{*} \sim 15$) requires $n_{\rm RS} \sim (1.5-3.0) \times 10^5$ cm$^{-3}$, depending on the composition. We estimate the cooling time behind the reverse shock as follows: 
\begin{eqnarray} 
t_{\rm cool} & \sim & \frac{3 kT_{\rm e} n_{\rm e} V}{L} \nonumber \\
& \sim & \left\{
\begin{array}{rl}
\frac{3 kT_{\rm e} M_{\rm RS}}{m_{\rm H} L} \sim 14,000 \ {\rm day} \ , & \ \mbox{for H-rich composition} \ ,\\
\frac{9 kT_{\rm e} M_{\rm RS}}{5 m_{\rm H} L} \sim 9,000 \ {\rm days} \ , & \ \mbox{for He-rich composition} \ .
\end{array}\right.
\end{eqnarray}
Here we set the temperature by the one from the low-temperature component (0.97 keV) to obtain the lower limit. Thus, we conclude that the reverse shock is adiabatic at 500 days. Then, we check the assumption of the electron-ion equipartition in deriving the electron temperature from the shock velocity:  
\begin{eqnarray}
t_{\rm e-i} & \sim & 29 \left(\frac{T_{\rm e}}{10^9 \ {\rm K}}\right)^{1.5} \left(\frac{n_{\rm e}}{10^8 \ {\rm cm}^{-3}}\right)^{-1} \ \ {\rm days}  \nonumber \\
& \sim & 40 - 100 \ {\rm days} \ . 
\end{eqnarray}
Here, the range corresponds to the different compositions and we adopt the temperature from the high-temperature component to provide the upper limit for the equipartition time scale. We conclude that the ion and electrons are largely in equipartition behind the reverse shock. 

The optical depth to X-rays by photoelectric absorption within the CSM is 
\begin{equation} 
\tau_{\rm CSM} \sim 0.012 \left(\frac{A_{*}}{15}\right) \left(\frac{R}{8 \times 10^{16} \ {\rm cm}}\right)^{-1} \ , 
\end{equation} 
therefore negligible at 500 days. This estimate assumes that the ionization state is largely neutral, therefore provides the upper limit. The column density in the interacting region at 500 days is (assuming the H-rich abundance), 
\begin{equation} 
N_{\rm RS} \sim \frac{M_{\rm RS}}{4 \pi R^2 m_{H}} \sim 1.9 \times 10^{20} \ {\rm cm}^{-2} \ .
\end{equation}
This is only comparable to the Galactic column density -- it would not produce detectable absorption as the temperature at the reverse shock, $\sim 1$keV, will lead to the high ionization state. The results (negligible absorptions both in CSM and the interacting region) are consistent with the result of the spectral fitting. Note that the column densities in these two regions are comparable and both evolve as a function of $R$ in the same way ($\propto 1/R$). Therefore, we expect that the larger contribution to absorption is always from the unshocked CSM (as the expected temperature is lower than behind the reverse shock), while the reverse shock contribution can be large (comparable to the CSM) only if/when the reverse shock is in the cooling regime. 

Through the above estimates, we conclude that the reverse shock is in an adiabatic regime, in electron-ion equipartition, and the absorption is negligible at 500 days. These justify a few assumptions we made in deriving the CSM density, $A_{*} \sim 15$. 

\section{Implications for The Early X-ray Emission}
A question is then what happened in the X-ray emission in the early phase. If we assume that the shock velocity follows the time dependence of $t^{-0.06}$ \citep[for $n = 20$; ][]{chevalier1982}, we expect that the temperature behind the reverse shock would have been $\sim 1.6$ keV (low-temperature component) and $\sim 4.2$ keV (high-temperature component) at 7 days, $\sim 1.5$ keV and $\sim 3.9$ keV at 12 days, and $\sim 1.3$ keV and $\sim 3.5$ keV at 33 days. Interestingly, the expected temperature is consistent with that inferred for the X-ray emission at 33 days (see \S 2). This indicates that a large fraction, if not all, of the X-ray luminosity at 33 days was indeed provided by the thermal emission from the reverse shock. 

Having determined the CSM density at $\sim 8 \times 10^{16}$ cm (corresponding to the mass loss at $\sim 1,300 (v_{\rm w}/20 \ {\rm km s}^{-1})^{-1}$ years before the SN explosion) from the data at $\sim 500$ days, the temporal evolution in the luminosity (therefore the mass loss history) is our next interest. The X-ray luminosity at the late phase ($\sim 500$ days) is smaller than that at $\sim 30$ days by a factor of $\sim 6$ (note that the difference is slightly larger than the simple comparison of the photon counts as the light curve is extracted from the spectral models). Here the X-ray luminosities in the early phase shown in Figure 6 are extracted from the power law fit, but we confirmed that extracting flux using the two-component thermal plasma model would not introduce large differences. The temporal behavior in these epochs is roughly $L (0.3-8 \ {\rm keV}) \propto t^{-0.65}$. 

The expected evolution is $L \propto \Lambda n_{\rm RS} M_{\rm RS}$. If we assume the bremsstrahlung ($\Lambda \propto T_{\rm e}^{0.5}$) and a steady-state mass loss ($s = 2$), the expression is reduced to a frequently used expression \citep{chevalier2003,chevalier2006, maeda2013a}, $L \propto V R^{-1} \propto t^{-1}$ (hereafter $R$ is the radius of the contact discontinuity and $V \equiv dR/dt$) independent from the outer ejecta density gradient ($n$). On the other hand, we expect that the temperature behind the reverse shock is fairly low from the beginning (see above), and thus the metal cooling is important -- then we can approximate the cooling function as $\Lambda \propto T_{\rm e}^{-0.67}$ \citep{chevalier2003}. Then, denoting the evolution of $R$ as $R \propto t^{m}$, we obtain $L \propto V^{-1.34} R^{-1} \propto t^{1.34 - 2.34 m}$. If we adopt $m \sim 0.94$ as expected for $(n, s) = (20, 2)$, then $L \propto t^{-0.86}$ -- this is steeper than the observed slope, and the expected luminosity at 33 days would exceed the observed luminosity by a factor of 1.8. We note however that $L$ is roughly scaled as $L \propto n_{\rm CSM}^2 \propto A_{*}^2$, and thus if there was a variation in either the mass loss rate or the wind velocity only by $\sim 30$\% over the time period between $\sim 1,300$ years to $\sim 100$ years before the explosion (for $v_{\rm w} \sim 20$ km s$^{-1}$), the expected evolution is consistent with the observed evolution between 33 days and $\sim 500$ days. 

More specifically, the expression for the evolution of the luminosity from the reverse shock can be generalized as follows: 
\begin{equation}
L \propto V^{-1.34} R^{(3-2s)} \propto t^{(1.66 - 2s)m + 1.34} \ . 
\end{equation}
As demonstration, if $s \sim 1.9$ as is just slightly flatter than the steady-state mass loss, then we obtain $L \propto t^{-0.67}$ for $n \sim 20$ as is consistent with the observation. The slope $s = 1.9$ corresponds to the decrease of the CSM density at $33$ days by $\sim 30$\% as compared to the steady-state case, as is expected from the above simple argument. In other words, the CSM density slope close to the steady-state wind can explain the X-ray evolution from 33 days to 500 days as the emission from the reverse shock. 

We check some conditions to justify the above mentioned situation. First, the cooling time scale is estimated as 
\begin{equation}
\frac{t_{\rm cool}}{t} \propto \frac{T_{\rm e} M_{\rm RS}}{L t} \propto t^{0.47} \ , 
\end{equation}
for the steady-state mass loss and $R \propto t^{0.94}$. Normalizing this by the estimate at 500 days ($t_{\rm cool}/t \sim 18 - 28$, depending on the composition), we obtain $t_{\rm cool}/t  \sim 2.3 - 3.5$ at 7 days, $\sim 3.1 - 4.9$ at 12 days, and $\sim 5.0 - 7.8$ at 33 days. Therefore, the reverse shock must be adiabatic at any epochs from the beginning. The electron-ion equipartition time scale is 
\begin{equation}
\frac{t_{\rm e-i}}{t} \propto \frac{T_{\rm e}^{1.5} n_{\rm e}^{-1}}{t} \propto t^{(5 m - 4)} \propto t^{0.7}\ . 
\end{equation}
Therefore, this ratio increases monotonically as a function of time, and with our estimate that it is in equipartition at 500 days, this is also satisfied in the earlier epochs. The equipartition is thus always archived behind the reverse shock, justifying our estimate of the temperature. While we assumed $s = 2$ in the above estimates, the results are not sensitive to it as long as $s$ is close to 2. 

The photoelectric absorption within the unshocked CSM, which is $\tau_{\rm CSM} \sim 0.012$ at 500 days, is $\tau_{\rm CSM} \propto R^{(1-s)} \propto t^{m(1-s)}$. Therefore, $\tau_{\rm CSM} \propto t^{^-0.94}$ for $s = 2$ or $t^{-0.85}$ for $s = 1.9$. This results in  $\tau_{\rm CSM} \sim 0.45 - 0.66$ at 7 days, $\sim 0.29 - 0.40$ at 12 days, and $\sim 0.12 - 0.15$ at 33 days. Thus the CSM could absorb $\sim 40$ \% of the emitted X-ray at 7 days, $\sim 30$\% at 12 days, while at 33 days it is at most 15\%. We note that the reverse shock would not make larger absorption than CSM under the condition relevant to SN 2011dh (see \S 3). These estimates are upper limit, since the CSM may be at a high ionization state due to the high energy photons from the interacting region. Thus, it is consistent with our fit results where we do not identify the evidence of huge absorption in addition to the Galactic one. These estimates, cooling time scale, equipartition time scale, absorption, show that all the assumptions in our derivation are justified from the early to the late phases.

Interpretation of the X-ray emission mechanism at $\lsim 10$ days is more complicated. The thermal X-ray luminosity from the reverse shock as extrapolated by our model constructed using the X-ray data at $\sim 500$ days and 33 days is lower than the observationally derived luminosity by a factor of $\sim 2$ at $\lsim 10$ days. Together with the spectral fit results for the spectra at 7 and 12 days (\S 2) which do not favor a thermal emission with properties (e.g., temperatures) predicted from the reverse shock model developed for the late-phase data, the X-ray emission at $\lsim 10$ days requires an additional mechanism. We note that, despite the possible dominance of this additional mechanism, about half of the X-ray luminosity at $\lsim 10$ days could indeed be contributed by the thermal emission from the reverse shock, therefore the interpretation of the early phase data by non-thermal mechanisms need to be revisited with this finding. 

What is the nature of this additional X-ray emission mechanism in the early phases at $\lsim 10$ days? Based on the spectral softening in the early phase (see also \S 2 and Figure 5), \cite{sasaki2012} argued that the early phase X-ray spectra are composed of two components, one persistent softer emission and the other decaying harder component. They suggested that the softer component is either the thermal emission from the reverse shock or the inverse Compton (IC) emission, while the harder one is likely the bremsstrahlung emission from the forward shock \citep[see also][]{campana2012}. With the present estimate of the non-negligible reverse shock component at these early phases, we suggest that the softer component is identified as the reverse shock component. 

For the origin of the hard component in the early phase, we argue that the IC mechanism is more likely than the bremsstrahlung emission from the forward shock. By identifying the reverse shock contribution, we can estimate the luminosity of the thermal emission from the forward shock (Fig. 6). If we extrapolate the CSM density structure as determined from the data between $33$ and $\sim 500$ days (i.e., $s \sim 1.9$), the forward shock contribution is far below the observed luminosity. One may argue that the mass loss rate just before the explosion might be enhanced by a large factor to increase the thermal emission behind the forward shock \citep{sasaki2012}, but our investigation of this possibility (see Appendix) suggests that this scenario is incompatible with the evolution of the radio emission and thus is quite unlikely. 

Another possibility is the IC scattering of the photospheric photons \citep[e.g., ][]{soderberg2012,maeda2012}. Thanks to large number densities both in the seed photons and relativistic electrons, the IC is suggested to be important in the early phase. A potential difficulty in this scenario is that the spectrum is very hard ($\Gamma \sim 1.2$ for the spectrum at 7 days)\footnote{We estimate that the Coulomb interaction cooling is not significant even at 7 days, for the parameters of SN 2011dh, thus this is not responsible to alter the electron spectrum.}, requiring a very hard spectrum of the corresponding non-thermal electrons at the forward shock ($N(E) \propto E^{-1.4}$). This is much flatter than the radio synchrotron-emitting electrons ($N(E) \propto E^{-3}$). In this respect, we note that \citet{maeda2012} suggested that the interpretation of the early-phase X-ray emission as the IC mechanism requires that the electrons at $\gamma \lsim 50$ (for the IC emission) should be an intrinsically different component than that at $\gamma \gsim 50$ (for the radio synchrotron emission), and the former component should have a cut off at a low energy (therefore a flat spectrum at a low energy). Therefore, while further study is required, the IC mechanism remains a possible (additional) mechanism for the early-phase X-ray emission.

\section{Conclusions and Discussion}

We have found a strong X-ray emission from SN IIb 2011dh at $\sim 500$ days after the explosion. Such a late-time X-ray observation is quite rare for extragalactic SNe -- especially among SE-SNe, it has been done only for SN IIb 1993J \citep{immler2001} and SN Ic 1994I \citep{immler2002}. We have shown the thermal nature of the late-time X-ray emission, which must come from the reverse shock. The reverse shock must be in an adiabatic phase, as evidenced by various arguments. Therefore, we can provide a solid measurement of the CSM density as $A_{*} \sim 15$, which translates into the mass loss rate in the final $\sim 1,300$ years before the explosion as $\sim 3 \times 10^{-6} M_{\odot} (v_{w} / 20 \ {\rm km} \ {\rm s}^{-1})$. We have checked various sources of uncertainties in our estimate, and conclude that our derivation of the CSM density and the mass loss rate are robust. 

The discovery of the YSG progenitor of SN IIb 2011dh makes the progenitor evolution of this object highly interesting. The derived mass loss rate should be an extremely strong constraint on any evolution models, either a single or binary \citep{georgy2012,benvenuto2013}. We note that this is a rare case where the mass loss rate of a SE-SN progenitor has been conclusively determined in a way free from assumptions in radio and X-ray emission mechanisms. 

This is the first report to clarify many properties of the SN ejecta and the CSM through X-ray emissions for the YSG progenitor. In this respect, we find (1) high density CSM, (2) existence of the outer envelope of the progenitor at least $\sim 0.013 M_{\odot}$, (3) steep density gradient in the outer envelope. These are not expected from a WR progenitor, strengthening the case of the YSG progenitor and suggesting quite different properties in the  progenitor/environment of a SE-SN from an extended progenitor as compared to a compact, WR-type star. The above properties are shared with a prototypical SN IIb 1993J. The CSM density is between `compact' SE-SNe \citep{chevalier2010} and SN 1993J with an RSG progenitor, linking the properties of the CSM along types of the progenitor star within SE-SNe. 

The steep outer density gradient ($n \sim 20$) is similar to what was found for SN IIb 1993J. A simple hydrodynamic model with either radiative or fully convective progenitor envelopes, $n \sim 10 - 12$, do not predict such steep density gradient. This will be a strong constraint on the nature of the envelope of the YSG (and RSG) progenitor. Also, we note that this finding is important in interpreting the very early UV/optical emission from SE-SNe with an extended progenitor. Frequently a fiducial value of $n \sim 10$ is assumed in analyzing such data \citep[e.g., ][]{rabinak2011}, but this would lead to significant errors if applied to SE-SNe from a progenitor with an extended envelope. 

The derived mass loss rate is indeed close to the `standard wind' mass-loss rate, in the final centuries toward the core-collapse, adopted in stellar evolution models of stars with $M_{\rm ZAMS} \sim 12 - 15 M_{\odot}$\citep{georgy2012}. Note that this mass range was well constrained for SN 2011dh through optical emission models \citep{bersten2012}. We emphasize that the CSM density distribution is close to that expected from the steady-state wind, with variation at the level of $\sim 30$\% within the last thousand years. Such a mass loss is unable to expel all the hydrogen-rich envelope within the life time of the progenitor star at the giant stage. Namely, our results suggest that in the final centuries the mass loss was just consistent with the YSG wind mass loss, and this steady mass loss cannot be the main driver of the whole hydrogen envelope stripping - this indicates that the mass loss rate was at some point much higher than that in the last thousand years before the explosion, supporting the binary interaction \citep{benvenuto2013} as the main driver of the mass loss to produce the YSG progenitor. 

Deriving the mass loss rate (and the CSM density) provides also a very strong constraint in understanding still-debated non-thermal/radiation physics in these wavelengths. With $A_{*} \sim 15$ as robustly determined, we can address the efficiency of the acceleration of non-thermal electrons producing the early-phase synchrotron radio emission as $\epsilon_{\rm e} \sim 0.01$ and the efficiency of the magnetic field amplification as $\epsilon_{B} \sim 5 \times 10^{-3}$ \citep{maeda2012}. 

By estimating the contribution of thermal emission from the reverse shock to the X-ray luminosity in the early phase, we conclude that it is very likely that the X-ray at 33 days was dominated by the same thermal component. The thermal emission from the reverse shock also accounts for about half of the X-ray luminosity at $\lsim 10$ days. Therefore, the soft persistent component in the early phase \citep{sasaki2012} should be identified as the reverse shock thermal emission. The early-phase X-ray spectra showed an additional hard component \citep{sasaki2012}. We find that it is unlikely that this component is the thermal emission (bremsstrahlung) from a forward shock -- such a solution requires a very large density jump by a factor of $\sim 15$ and this is inconsistent with radio data (see Appendix). An interesting alternative is the IC scattering of the SN photospheric photons \citep{bjornsson2004,soderberg2012,maeda2012}, and in this case the non-thermal electrons at low energy ($\gamma \lsim 50$) should have a flatter spectrum than those at higher energy. This may be consistent with the earlier suggestion that the IC scenario requires that these low energy electrons are a different component than the radio-synchrotron emitting electrons \citep{maeda2012}. While we suggest the IC scenario is more likely than the thermal emission from the forward shock, more detailed study will be necessary to resolve this issue. 

The shock wave has expanded to $\sim 8 \times 10^{16}$ cm at $\sim 500$ days and the swept ejecta mass is $M_{\rm RS} \sim 0.013 M_{\odot}$, which is either H-rich or He-rich. This indicates that the reverse shock should be still in the outer envelope of the progenitor star. Namely, we constrain the mass of the envelope in the YSG progenitor of SN IIb 2011dh to be $\gsim 0.013 M_{\odot}$. We note this is fully consistent with the early-phase optical spectral modeling, from which the hydrogen mass of $\sim 0.024 M_{\odot}$ in the outer layer has been derived \citep{arcavi2011}. 

Our estimate of the CSM density is not sensitive to particular interpretation. For example, we have shown that the derived CSM density is insensitive to the composition in the outer envelope. We also checked this issue, by performing the same analyses based on the one-component sub-solar model. There we have confirmed that all the arguments presented with the two-component model are qualitatively reproduced by the sub-solar model as well. The only difference, in a quantitative sense, is that the required CSM density is reduced by a factor of $\sim 2$. This stems from overall low temperature in the sub-solar model, lacking the high temperature component, leading to a larger cooling rate than in the two-component model. While we regard such a low metallicity model unlikely, we note that once an independent estimate of the metallicity is given, that will discriminate these two scenarios. In any case, this would not alter any of our conclusions. 

Finally, we note that, while our analyses are based on the assumption that the CSM density is represented by a smooth distribution ($\rho_{\rm CSM} \propto r^{-s}$), most of our conclusions are not influenced by this assumption. First of all, the derived relation between the density and the mass of the emitting region (i.e., equation 5) is basically independent from this assumption. Since the reverse shock position within the ejecta outer envelope is mostly determined by the CSM mass already encountered by the forward shock, the mass loss rate derived by this study can be regarded as the {\em average} mass loss in the final 1,300 years in the case with the huge CSM density variation (assuming $v_{\rm w} = 20$ km s$^{-1}$). Interpreting the X-ray emission as a thermal emission from the {\em forward} shock at the putative dense CSM will lead to a similar conclusion, since the density behind the forward shock is required to be similar to the reverse shock case, as constrained by the observed characteristic energy scale in the X-ray spectrum. Moreover, we emphasize that the interpretation based on the smooth CSM density distribution provides consistent results to various observed behaviors, and indeed the derived distribution is close to the steady-state mass loss. Therefore, it does not require the dense CSM shell, and as such we suggest that the smooth CSM distribution is very likely the case. 

\acknowledgements 
The scientific results reported in this article are based on data obtained from the Chandra Data Archive. K.M. thanks Yoshihiro Ueda for his help and advises on the {\bf XSPEC} spectral fittings. The work by K.M. is partly supported by World Premier International Research Center Initiative (WPI Initiative), MEXT, Japan. The authors acknowledge financial support by Grant-in-Aid for Scientific Research (No. 23740141 for K.M., 25800119 for S.K., 
22684012 for A. B., and 23340055 for Y. T) from the Japanese Ministry of Education, Culture, Sports, Science and Technology (MEXT). S.K. is supported by the Special Postdoctoral Researchers Program in RIKEN.

\appendix

\section{A Possibility of Bremsstrahlung Emission from the Forward Shock for Emission at $\lsim 10$ Days}

\begin{figure}
\begin{center}
        \begin{minipage}[]{0.45\textwidth}
                \epsscale{1.1}
                \plotone{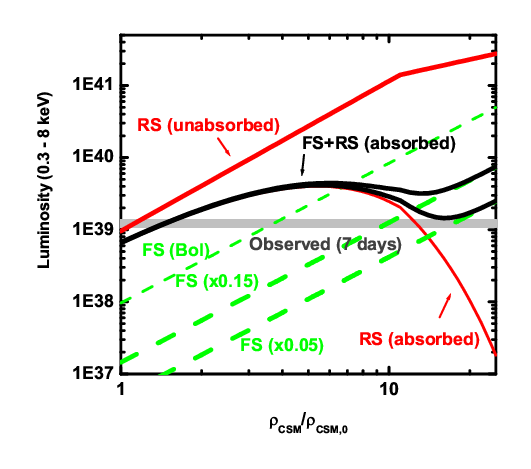}
        \end{minipage}
\end{center}
\caption
{Expected properties of the thermal X-ray emission from SN IIb 2011dh at 7 days, as a function of the CSM density normalized by the extrapolation from the late-phase X-ray behavior. For our standard case ($\rho_{\rm CSM}/\rho_{{\rm CSM}, 0} = 1$), the forward shock contribution is negligible, and the reverse shock contribution accounts for about a half of the observed X-ray luminosity. The unabsorbed reverse shock contribution is shown by a thick red line, where the break at $\rho_{\rm CSM}/\rho_{{\rm CSM}, 0} \sim 10$ is due to a transition from an adiabatic to a cooling regime. The possible effect of the cooling shell in creating additional absorption is shown by the thin-red line, which assumes the maximally allowed absorption in the interaction region (i.e., neutral compositions due to the cooling) and provides the lower limit for the reverse shock X-ray luminosity. The forward shock contribution is shown by the dashed green lines, where the thin line is for the bolometric luminosity while the two thick lines are for the fraction of 0.05 and 0.15 emitted in the 0.3 - 8 keV range (as appropriate for $kT_{\rm FS} \sim 50 - 150$ keV). The only solution to account for the observed property by the forward shock thermal emission is found for $kT_{\rm FS} \gsim 150$ keV and $\rho_{\rm CSM}/\rho_{{\rm CSM}, 0} \sim 15 - 20$.  
\label{fig9}}
\end{figure}

While we argued in \S 4 that the bremsstrahlung emission from the forward shock does not explain the hard component in the early-phase ($\lsim 10$ days), this is based on the assumption that the CSM density distribution is smooth. Since the dense CSM shell encountered before $\lsim 10$ days was (qualitatively) suggested as an alternative scenario \citep{sasaki2012}, it is interesting to quantitatively investigate if such a scenario is consistent with the observations. In this section, we show that this scenario can indeed work as long as only the X-ray emission is concerned, but it creates a tension to the radio evolution. 

Figure A.1 shows the expected thermal emission luminosity, both for the forward shock and reverse shock, if one increases the CSM density (at 7 days; $\rho_{\rm CSM}$) as compared to the extrapolation from the later phases ($\rho_{{\rm CSM}, 0}$). Here, we calculate the X-ray luminosity at a given epoch, adopting the scaling relations as follows: 
\begin{eqnarray}
& L_{\rm FS} & \propto \rho_{\rm CSM}^2 T_{\rm FS}^{0.5} \propto \rho_{\rm CSM}^{1.94} \ , \\
& L_{\rm RS (unabsorbed)} & \propto \rho_{\rm CSM}^2 T_{\rm RS}^{-0.67} \propto \rho_{\rm CSM}^{2.08} \ , \\
& \tau_{\rm RS (maximum)} & \propto \rho_{\rm CSM} \ . 
\end{eqnarray}
Here we have assumed the density slope of $n \sim 20$ to compute the scaling for the temperature, but anyway this produces just a minor correction to the expected behavior for the luminosity (i.e., $\propto \rho_{\rm CSM}^2$). As the CSM density is (hypothetically) increased, {\em both} the forward shock and reverse shock contributions increase. For the high density the reverse shock is in the cooling regime and the photons may be absorbed in the interaction region (see \S 4 for discussion on the cooling and the absorption), that will apply for large $\rho_{\rm CSM}/\rho_{{\rm CSM}, 0}$. Considering the maximally allowed absorption (assuming the interacting region is neutral in ionization), we find that there is a solution to explain the X-ray luminosity at 7 days by the bremsstrahlung emission from the forward shock. The conditions are $\rho_{\rm CSM}/\rho_{{\rm CSM}, 0} \sim 15 - 20$ and $T_{\rm FS} \gsim 150$ keV. This corresponds to a local CSM density (as defined by $\rho_{\rm CSM} \equiv 5 \times 10^{11} A_{*} r ({\rm 7 days})^2$) of $A_{*} \sim 150 - 200$, taking into account the CSM density slope of $s = 1.9$.

Therefore, it is in principle possible to explain (the hard component of) the early-phase {\em X-ray property} by the bremsstrahlung emission from the interacting region created by the dense CSM shell, as qualitatively suggested by \citet{sasaki2012}. However, we find that it introduces a discrepancy in the radio data. The radio light curves exhibited a smooth behavior \citep{soderberg2012,krauss2012,horesh2013}, which is difficult to reconcile with such a huge density jump by a factor of $\sim 15$ in a short period of time. Figure A. 2 shows the radio spectra of SN 2011dh at 4 and 17 days after the explosion. Taking our fiducial model ($\rho_{\rm CSM}/\rho_{{\rm CSM}, 0} = 1$, i.e., $A_{*} = 15$ as determined at $\sim $ 500 days), we can explain the general behavior of the radio spectral evolution without fine tuning. However, increasing the density by a factor of $15$ (only at $\lsim 10$ days) results in totally different spectrum at 4 days -- too high synchrotron self-absorption frequency and too large flux density at high frequencies. 

\begin{figure}
\begin{center}
        \begin{minipage}[]{0.45\textwidth}
                \epsscale{1.1}
                \plotone{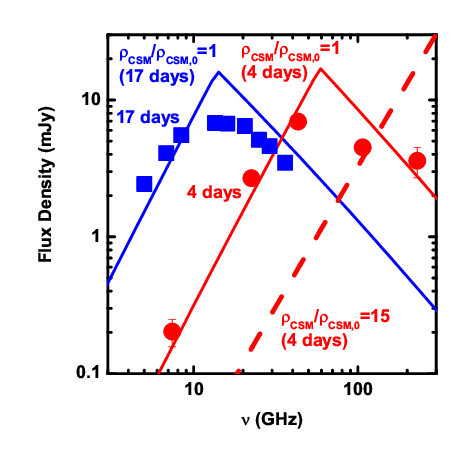}
        \end{minipage}
\end{center}
\caption
{Radio synchrotron emission from SN IIb 2011dh at 4 days (red) and 17 days (blue). The expectation from our fiducial model (solid lines) and the one with the density enhancement by a factor of 15 at 4 days since the explosion (dashed line) are shown. The radio data are from \citet{soderberg2012} and \citet{horesh2013}. Note that the radio-synchrotron model \citep[developed for the optically thin and thick limits; see ][]{maeda2012,maeda2013a} uses a simple prescription for the intermediate optical depth to the synchrotron self-absorption (SSA), thus the detailed shape around the SSA peak should not be compared directly to the observational data. Also, the model here without a fine-tuning is not intending to provide a detailed fit. 
\label{fig10}}
\end{figure}





\begin{thebibliography}{}

\bibitem[Arcavi et al.(2011)]{arcavi2011}
Arcavi, I., et al. 2011, ApJ, 742, L18

\bibitem[Arnaud(1996)]{arnaud1996}
Arnaud, K.A. 1996, in ASP Conf. Ser. 101, `Astronomical Data Analysis Software and Systems V.', ed. G.H. Jacoby \& J. Barnes (San Francisco: ASP), 17

\bibitem[Benvenuto et al.(2013)]{benvenuto2013}
Benvenuto, O., Bersten, C.M., Nomoto, K. 2013, ApJ, 762, 74

\bibitem[Bersten et al.(2012)]{bersten2012}
Bersten, C.M., et al. 2012, ApJ, 757, 31

\bibitem[Bufano et al.(2014)]{bufano2014}
Bufano, F., et al. 2014, MNRAS, in press (arXiv:1401.2368)

\bibitem[Bietenholz et al.(2012)]{bietenholz2012}
Bietenholz, M. F., Brunthaler, A., Soderberg, A. M., Krauss, M., Zauderer, B., Bartel, N., Chomiuk, L., Rupen, M. P. 2012, ApJ, 751, 125

\bibitem[Bj\"ornsson \& Fransson(2004)]{bjornsson2004}
Bj\"ornsson, C.-I., \& Fransson, C. 2004, ApJ, 605, 823 

\bibitem[Bresolin et al.(2004)]{bresolin2004}
Bresolin, F., Garnett, D.R., Kennicutt, R.C., Jr. 2004, ApJ, 615, 228

\bibitem[Campana \& Immler(2012)]{campana2012}
Campana, S., \& Immler, S. 2012, MNRAS, 427, L70

\bibitem[Cash(1979)]{cash1979}
Cash, W. 1979, ApJ, 228, 939

\bibitem[Chakraborti et al.(2013)]{chakraborti2013}
Chakraborti, S., et al. 2013, ApJ, 774, 30

\bibitem[Chevalier(1982)]{chevalier1982}
Chevalier, R.A., 1982, ApJ, 258, 790 

\bibitem[Chevalier \& Fransson(2003)]{chevalier2003}
Chevalier, R.A., \& Fransson, C. 2003, in `Supernovae and Gamma-Ray Bursters', Lecture Notes in Physics (Springer; New York), 171

\bibitem[Chevalier \& Fransson(2006)]{chevalier2006}
Chevalier, R.A., \& Fransson, C. 2006, ApJ, 651, 381 

\bibitem[Chevalier \& Soderberg(2010)]{chevalier2010}
Chevalier, R.A., \& Soderberg, A.M. 2010, ApJ, 711, L40 

\bibitem[Filippenko(1997)]{filippenko1997}
Filippenko, A.V. 1997, ARAA, 35, 309

\bibitem[Fransson, Lundqvist, \& Chevalier(1996)]{fransson1996}
Fransson, C., Lundqvist, P., \& Chevalier, R.A. 1996, ApJ, 461, 993

\bibitem[Georgy(2012)]{georgy2012}
Georgy, C. 2012, A\&A, 538, L8

\bibitem[Horesh et al.(2013)]{horesh2013}
Horesh, A., et al. 2013, MNRAS, 436, 1258 

\bibitem[Immler, Aschenbach, \&Wang(2001)]{immler2001}
Immler, S., Aschenbach, B., \& Wang, Q.D. 2001, ApJ, 561, 107

\bibitem[Immler et al.(2002)]{immler2002}
Immler, S., Wilson, A.S., \& Terashima, Y. 2002, ApJ, 573, L27

\bibitem[Kalberla et al.(2005)]{kalberla2005}
Kalberla, P.M.W., et al. 2005, A\&A, 440, 775

\bibitem[Krauss et al.(2012)]{krauss2012}
Krauss, M.I., et al. 2012, ApJ, 750, 40

\bibitem[Lodders(2003)]{lodders2003}
Lodders, K. 2003, ApJ, 591, 1220

\bibitem[Maeda(2012)]{maeda2012}
Maeda, K. 2012, ApJ, 758, 81

\bibitem[Maeda(2013a)]{maeda2013a}
Maeda, K. 2013a, ApJ, 762, L24

\bibitem[Maeda(2013b)]{maeda2013b}
Maeda, K. 2013b, ApJ, 762, 14

\bibitem[Maund et al.(2011)]{maund2011}
Maund, J.R., et al. 2011, ApJ, 739, L37

\bibitem[Nomoto et al.(1993)]{nomoto1993}
Nomoto, K., Suzuki, T., Shigeyama, T., Kumagai, S., Yamaoka, H., Saio, H. 1993, Nature, 364, 507

\bibitem[Nymark et al.(2009)]{nymark2009}
Nymark, T.K., Chandra, P., \& Fransson, C. 2009, A\&A, 494, 179

\bibitem[Rabinak \& Waxman(2011)]{rabinak2011}
Rabinak, I., \& Waxman, E. 2011, ApJ, 728, 63

\bibitem[Sahu et al.(2013)]{sahu2013}
Sahu, D.K., Anupama, G.C., Chakradhari, N.K. 2013, MNRAS, 433, 2

\bibitem[Sasaki \& Ducci(2012)]{sasaki2012}
Sasaki, M., \& Ducci, L. 2012, A\&A, 546, 80

\bibitem[Smith et al.(2001)]{smith2001}
Smith, R.K., Brickhouse, N.S., Liedahl, D.A., \& Raymond, J.C. 2001, ApJ, 556, L91

\bibitem[Smith et al.(2007)]{smith2007}
Smith, I.A., Ryder, S.D., B\"ottcher, M., Tingay, S.J., Stacy, A., Pakull, M., Liang, E.P. 2007, ApJ, 669, 1130

\bibitem[Soderberg et al.(2012)]{soderberg2012}
Soderberg, A.M., et al. 2012, ApJ, 752, 78 

\bibitem[Suzuki \& Nomoto(1995)]{suzuki1995}
Suzuki, T., \& Nomoto, K. 1995, ApJ, 455, 658

\bibitem[Van Dyk et al.(2011)]{vandyk2011}
Van Dyk, S.D., et al. 2011, ApJ, 741, L28

\bibitem[Van Dyk et al.(2013)]{vandyk2013}
Van Dyk, S.D., et al. 2013, ApJ, 772, L32

\bibitem[Woosley et al.(1994)]{woosley1994}
Woosley, S.E., Eastman, R.G., Weaver, T.A., \& Pinto, P.A. 1994, ApJ, 429, 300 

\end{thebibliography}
\end{document}